\documentclass[english,a4paper]{article}
\usepackage[T1]{fontenc}
\usepackage[cp1250]{inputenc}
\usepackage{amssymb}
\usepackage{hyperref}
\usepackage{amsfonts}
\usepackage{graphicx}
\usepackage{revsymb}
\usepackage{dcolumn}
\usepackage{bm}

\usepackage{color}
\usepackage{ulem} 

\newtheorem {theo} {Theorem} [section]

\newtheorem{rem}[theo]{Remark}

\newcommand{\arctanh}{\textrm{arctanh}}
\newcommand{\cn}{\textrm{cn}}
\newcommand{\sn}{\textrm{sn}}
\newcommand{\dn}{\textrm{dn}}
\newcommand{\mod}{\textrm{mod}}

\def\openone{\leavevmode\hbox{\small1\kern-3.3pt\normalsize1}}

\begin{document}
\title{The tennis racket effect in a three-dimensional rigid body}
\author{L. Van Damme\footnote{Laboratoire Interdisciplinaire
Carnot de Bourgogne (ICB), UMR 5209 CNRS-Universit\'e Bourgogne Franche Comt\'e, 9 Av. A. Savary, BP 47 870, F-21078 DIJON Cedex,
France}, P. Mardesic\footnote{Institut de Math\'ematiques de Bourgogne, UMR 5584 CNRS-Universit\'e Bourgogne Franche-Comt\'e, 9 Av. A. Savary, BP 47870 21078 Dijon Cedex, France}, D. Sugny\footnote{Laboratoire Interdisciplinaire
Carnot de Bourgogne (ICB), UMR 5209 CNRS-Universit\'e Bourgogne Franche Comt\'e, 9 Av. A. Savary, BP 47 870, F-21078 DIJON Cedex,
France}}

\maketitle

\begin{abstract}
We propose a complete theoretical description of the tennis racket effect, which occurs in the free rotation of a three-dimensional rigid body. This effect is characterized by a flip ($\pi$- rotation) of the head of the racket when a full ($2\pi$) rotation around the unstable inertia axis is considered. We describe the asymptotics of the phenomenon and conclude about the robustness of this effect with respect to the values of the moments of inertia and the initial conditions of the dynamics. This shows the generality of this geometric property which can be found in a variety of rigid bodies. A simple analytical formula is derived to estimate the twisting effect in the general case. Different examples are discussed.
\end{abstract}




\section{Introduction}
Understanding the dynamics of classical Hamiltonian systems remains a crucial goal with a variety of applications which go well beyond its mathematical description \cite{Golstein50,efstathiou}. In the case of integrable systems with few degrees of freedom \cite{landau,arnold}, an efficient approach is based on a geometric analysis to characterize the dynamical properties of the mechanical system \cite{cushman,fomenko}. Such geometric phenomena are typically at the origin of the robustness of particular effects which can be observed experimentally. Some examples, which are by now mathematically well understood, are given by the Hannay angle \cite{hannay}, the Hamiltonian monodromy \cite{Duistermaat}, or the Montgomery phase \cite{montgomery,levi,natario}. Another property is the Tennis Racket Effect (TRE), which occurs in the free rotational dynamics of a three-dimensional rigid body. Its name comes from the fact that it can be easily observed in a standard tennis racket. In spite of the apparently well-known character of TRE, this property is not discussed in standard text-books of classical mechanics \cite{Golstein50,landau,arnold}. To the best of our knowledge, the only mathematical study of this geometric effect has been made in~\cite{MSA91} and reproduced in \cite{cushman}. However, this work only focuses on a specific case where the moments of inertia satisfy a constraint of the form $I_1+I_2\simeq I_3$. The asymptotics and the robustness of the flip with respect to the values of the moments of inertia of the rigid body are not investigated. Note that the Montgomery phase also occurs in the free rotation of a rigid body when the angular momentum undergoes a cyclic trajectory. The goal of this paper is therefore to extend the preliminary analysis made in \cite{cushman,MSA91} in different directions. We choose a suitable set of Euler angles which allows us to give a complete and unified analytical and geometrical description of TRE. A simple asymptotic formula is derived to evaluate the value of the flip for any three-dimensional rigid body.

The outline of the paper is as follows. In Sec. \ref{sec2}, we introduce the model system and the different frames used to describe the rotational dynamics of the rigid body. A suitable set of Euler angles is proposed to globally parameterize the TRE. Sections \ref{sec3} and \ref{sect_euler_eq} are dedicated to the derivation of the Euler equations and to the geometric description of the corresponding solutions. In Sec. \ref{sec4}, we begin the study of the TRE by integrating the Euler equations. An asymptotic analytical formula is derived to estimate this effect for any rigid body in Sec. \ref{sect_gamma0}. This asymptotic analysis allows one to describe the robustness of the flip with respect to the shape of the body and to the initial conditions of the dynamics. Numerical  computations are proposed in Sec. \ref{sec5} to show the wide range of applicability of this effect. We compare the numerical results to the ones given by the analytical formula, and show that both coincide for a large class of initial conditions and rigid bodies. The flip is estimated for different tennis rackets. Conclusion and prospective views are given in Sec. \ref{sec6}. Possible extensions to quantum systems are discussed. Some technical computations are reported to~\ref{app}.
\section{The model system}\label{sec2}
\begin{figure}[h!]
\centering
\includegraphics[scale=0.6,trim=0 80 0 0,clip]{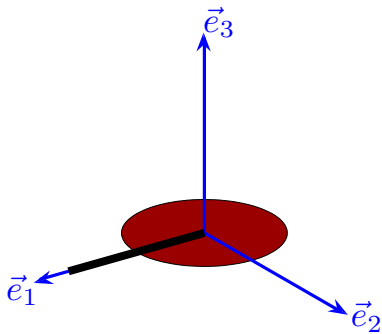}
\caption{(Color online) Representation of the frame attached to the tennis racket. The handle and the head of the racket are represented respectively by a thick black line and by a red disk.\label{fig_racket}}
\end{figure}

This paragraph is aimed at mathematically defining the TRE. We introduce the frame $(\vec{e}_1,\vec{e}_2,\vec{e}_3)$ attached to the racket~\cite{MSA91}, for which each axis is parallel to one of the inertia axes of the racket. We recall that the principal inertia axis $\vec{e}_1$ is parallel to the handle of the racket, the axis $\vec{e}_2$ is along the intermediate inertia axis, which lies in the plane of the head of the racket and is perpendicular to the handle, and the axis $\vec{e}_3$ is perpendicular to the plane of the head of the racket. The origin of the frame is the center of mass of the racket. This frame is displayed in Fig. \ref{fig_racket}. The experiment showing the TRE consists in throwing the racket in the air in order to make a rotation about one of the three inertia axes. It is well known that if the racket rotates around the $\vec{e}_1$- or $\vec{e}_3$- axis then the rotation is stable, while the motion is unstable around the $\vec{e}_2$- axis. This behavior can be observed for any three-dimensional rigid body \cite{Golstein50}. In the case of a tennis racket, the experiment shows that a flip of the head occurs with the unstable rotation. This means that after a full rotation of $2\pi$ of the handle in the air, the racket makes a rotation of approximately $\pi$ about its handle. This effect is known as the  \textit{tennis racket effect}. A video is available in the supplementary material to illustrate this effect~\cite{video}.

\begin{figure}[h!]
\centering
\includegraphics[scale=0.7,trim=80 60 70 40,clip]{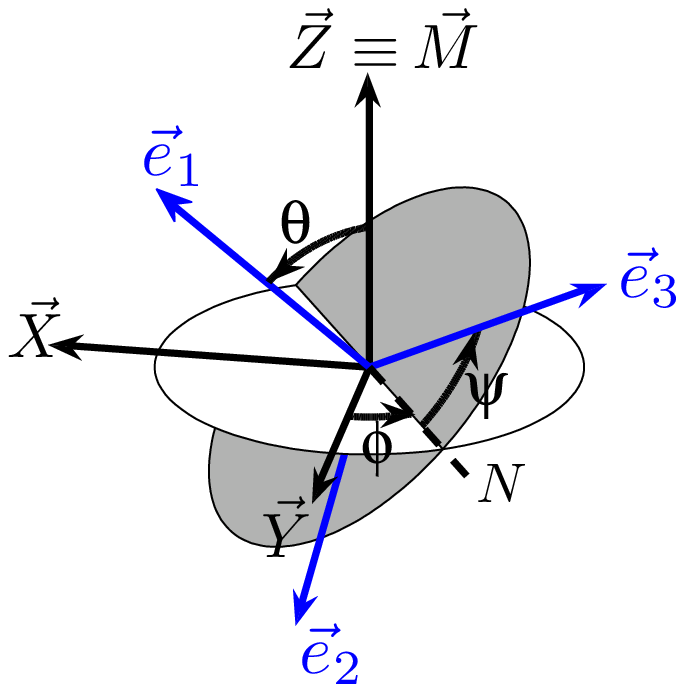}
\includegraphics[scale=0.7,trim=80 60 70 40,clip]{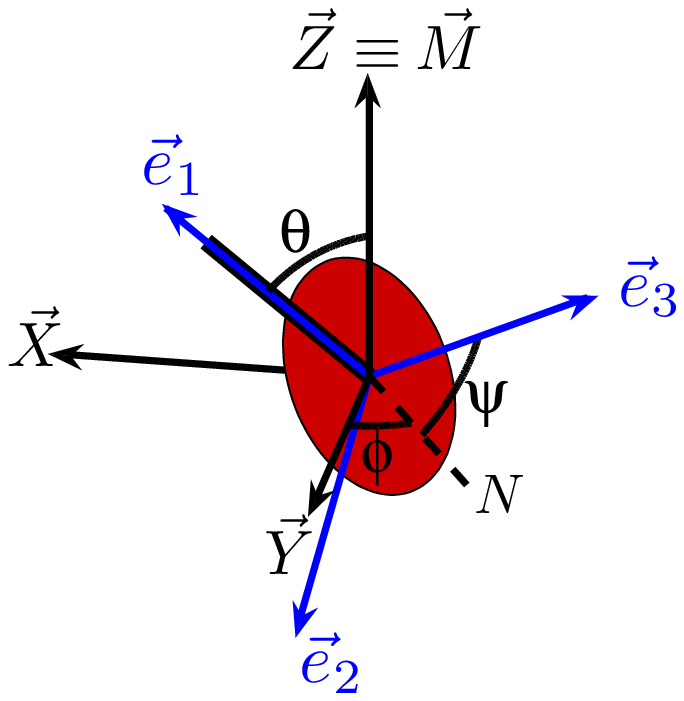}
\caption{(Color online) \textit{Left panel:} Standard representation of the Euler Angles. The angle $\phi$ describes the precession of the axis along $\vec{e}_1$ about the $\vec{Z}$- axis \ ($\equiv \vec{M}$). Note that $\phi$ is also the angle between $\vec{X}$ and the projection of $\vec{e}_1$ onto the $(\vec{X},\vec{Y})$-plane (see the attached video). The angle $\psi$ is the flip angle, and $N$ is the line of nodes. \textit{Right panel:} Representation of Euler angles with respect to the frame attached to the racket. In the racket experiment, the initial conditions are $\theta_0\simeq \pi/2$, $\psi_0\simeq 0$ and $\phi_0=0$ (see the text for details).\label{fig_angles}}
\end{figure}

The rotational dynamics of a three-dimensional rigid body about its center of mass is governed by Euler equations \cite{landau}, which are expressed in terms of Euler angles. We have chosen a set of Euler angles so as to describe easily the motion of the handle and the flip of the racket. We are interested in the motion of the moving frame $(\vec{e}_1,\vec{e}_2,\vec{e}_3)$ with respect to a static one, the laboratory frame, namely $(\vec{X},\vec{Y},\vec{Z})$. The origin of the two frames coincides. The total angular momentum $\vec{M}$ of the rigid body is conserved since no external moment is exerted on the system. The components $M_i$ of $\vec{M}$ in the $(\vec{e}_1,\vec{e}_2,\vec{e}_3)$- frame can be expressed in terms of the components $\Omega_i$ of the angular velocity vector $\vec{\Omega}$ as follows:
\begin{equation}
M_1=I_1\Omega_1,\quad M_2=I_2\Omega_2,\quad M_3=I_3\Omega_3,
\label{eq_omega}
\end{equation}
where $I_1$, $I_2$ and $I_3$ are the principal moments of inertia of the solid. For the stable rotation about the handle, the only non-zero component of $\vec{\Omega}$ is $\Omega_1$ and $\vec{M}$ is parallel to the axis $\vec{e}_1$. The vector $\vec{M}$ is parallel to $\vec{e}_3$ for the second stable rotation and to $\vec{e}_2$ for the unstable one. This latter corresponds to the case of the TRE. The dynamics of the moving frame can be described by the coordinates of its center of mass and three Euler angles $\phi$, $\theta$ and $\psi$~\cite{landau}. There is a natural way to define one of the axes of the laboratory frame by using the fact that the angular momentum $\vec{M}$ is a constant of the motion.
Following Ref.~\cite{landau}, we fix the $\vec{Z}$- axis of the laboratory frame along $\vec{M}$. This physical property allows us to define two Euler angles $\theta$ and $\psi$. The position of the two other axes along $\vec{X}$ and $\vec{Y}$ is given by the third Euler angle $\phi$. The two frames and the Euler angles are displayed in Fig.~\ref{fig_angles}.

With this choice of coordinates, the angle $\phi$ describes the rotation of the handle in space, since it corresponds to the precession of the handle about the angular momentum. The flip is given by the angle $\psi$ which is associated with the rotation about the handle of the racket represented by $\vec{e}_1$. As explained above, the TRE manifests for a rotation about $\vec{e}_2$, so that initially $\vec{e}_2$ is approximately collinear to $\vec{M}$. This corresponds to initial conditions of the form:
\begin{equation}
\theta_0\simeq\frac{\pi}{2},\quad\psi_0\simeq 0.
\end{equation}
Since $\vec{X}$ and $\vec{Y}$ are arbitrary directions, we can set $\phi_0=0$. Thus, in this system of coordinates, the TRE is defined as:
\begin{equation}
\Delta\phi=2\pi,\quad \Delta\psi\simeq\pi.
\end{equation}
Note that the value of the $\theta$- angle is not taken into account in this definition. We will see below for the TRE that if $\Delta\psi\simeq\pi$ then $\Delta\theta\simeq 0$.


\section{Dynamics of the angular momentum}\label{sec3}
\begin{figure}[h!]
\centering
\includegraphics[scale=0.45,trim=50 10 10 0,clip]{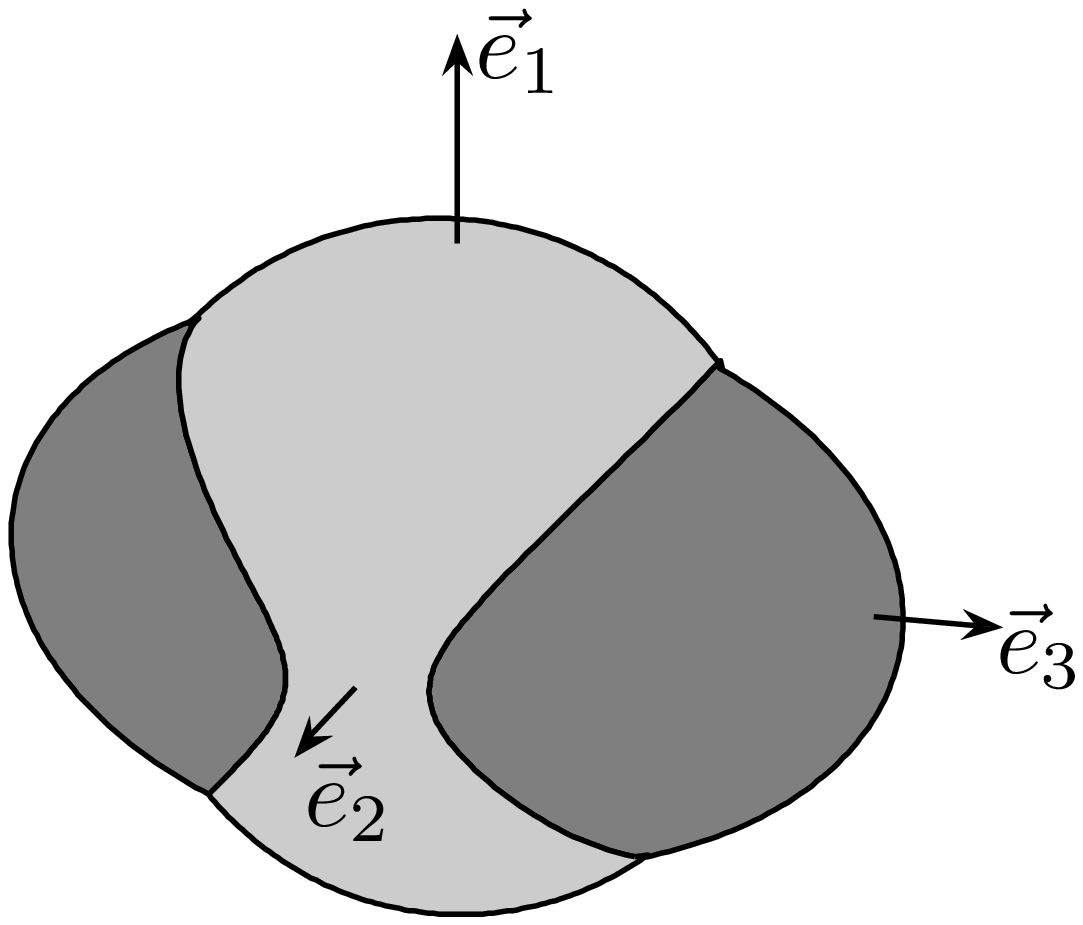}
\includegraphics[scale=0.45,trim=50 10 10 0,clip]{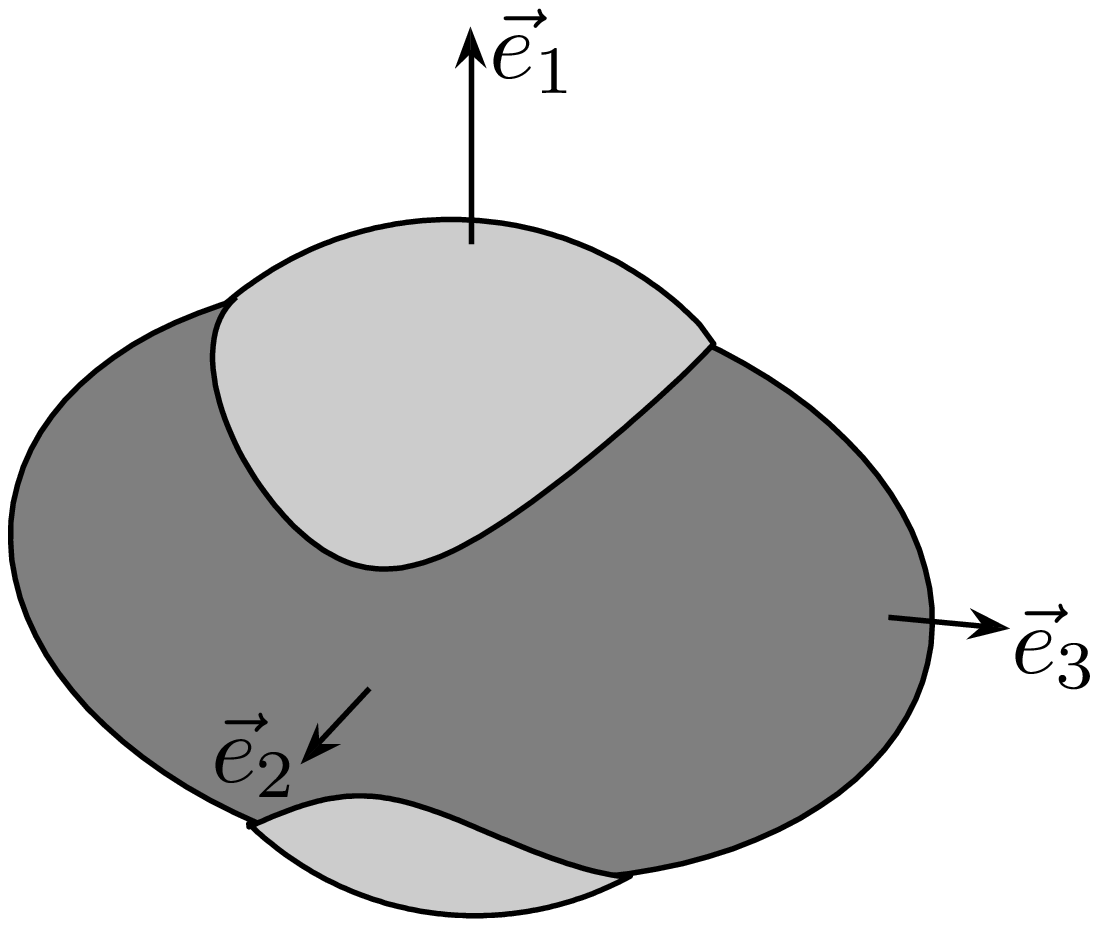}
\includegraphics[scale=0.5,trim=70 80 30 10,clip]{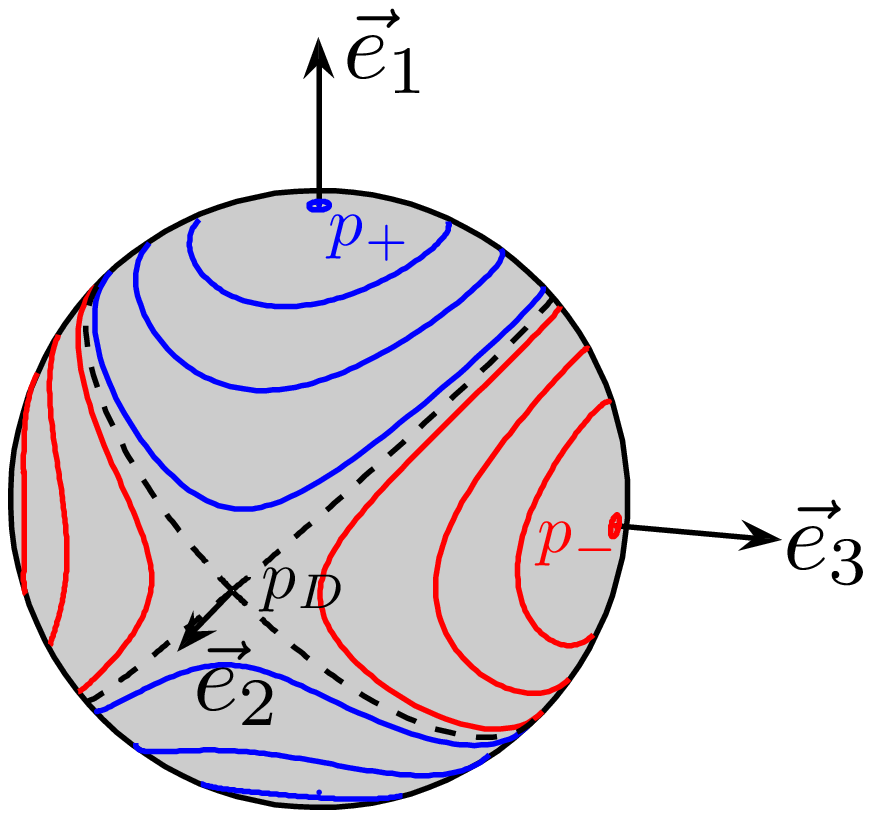}
\includegraphics[scale=0.5,trim=70 80 30 10,clip]{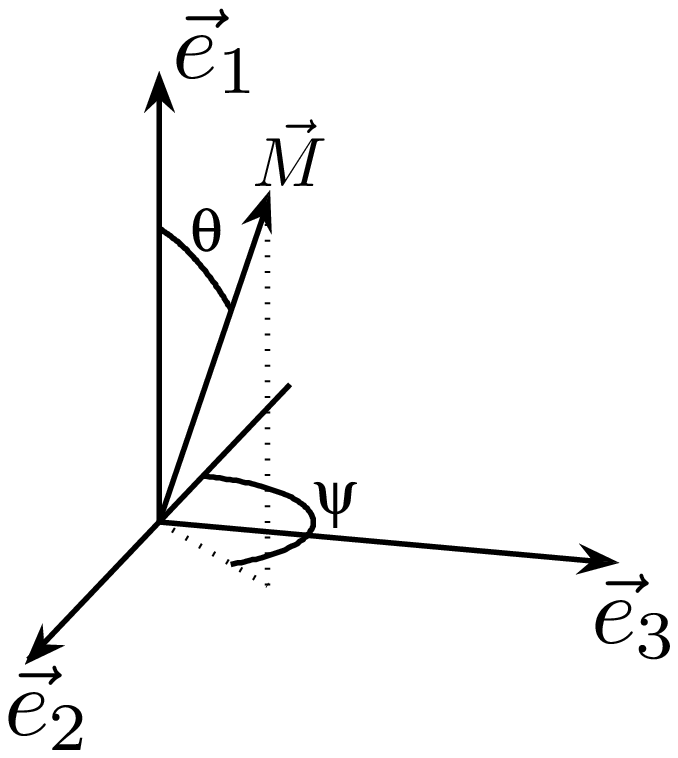}
\caption{(Color online) \textit{Upper panels:} Plot of the energy ellipsoid (dark-grey) and of the angular momentum sphere (light-grey) in the oscillating $2I_2E<M^2<2I_3E$ (left) and in the rotating cases $2I_1E<M^2<2I_2E$ (right). The solution $\vec{M}(t)$ of Eq.~(\ref{eq_dMi}) belongs to the intersection of the two surfaces. \textit{Lower left panel:} Plot of the solution $\vec{M}(t)$ of Eq.~(\ref{eq_dMi}) on the angular momentum sphere for different values of $E$. The rotating solutions are represented in blue (dark grey) and the oscillating ones in red (light grey). The dotted line depicts the separatrix, given by $2I_2E=M^2$. The two points $p_+$ and $p_-$ are the stable equilibrium points, while $p_D$ is the unstable one. \textit{Lower right panel:} Representation of the two Euler angles $\theta$ and $\psi$ in the frame $(\vec{e}_1,\vec{e}_2,\vec{e}_3)$. \label{fig_sphere}}
\end{figure}

Euler equations describe the dynamics of the coordinates of the angular momentum vector $\vec{M}$ in the rotating frame attached to the body. In the rest of the paper, we will assume that $0\leq I_1\leq I_2\leq I_3$. The dynamics is governed by the set of equations~\cite{Golstein50,landau}:
\begin{equation}\label{eq_dMi}
\begin{array}{lll}
&\frac{dM_1}{dt}=-\left(\frac{1}{I_2}-\frac{1}{I_3}\right)M_2M_3, \\
&\frac{dM_2}{dt}=\left(\frac{1}{I_1}-\frac{1}{I_3} \right)M_1M_3, \\
&\frac{dM_3}{dt}=-\left(\frac{1}{I_1}-\frac{1}{I_2} \right)M_1M_2.
\end{array}
\end{equation}
The solutions of the differential system are given by Jacobi's elliptic functions $\cn$, $\sn$ and $\dn$ \cite{abramowitz}. In this work, we
use a geometric approach presented in~\cite{Golstein50,landau} to describe the motion of $\vec{M}$. The system admits two constants of the motion, the energy and the angular momentum:
\begin{equation}
2E=\frac{M_1^2}{I_1}+\frac{M_2^2}{I_2}+\frac{M_3^2}{I_3},\quad M^2=M_1^2+M_2^2+M_3^2.
\label{eq_first_int}
\end{equation}
In the rotating frame $(\vec{e}_1,\vec{e}_2,\vec{e}_3)$, the two first integrals are represented by two surfaces, namely the energy ellipsoid, of radii $\sqrt{2I_1E}$, $\sqrt{2I_2E}$ and $\sqrt{2I_3E}$, and the angular momentum sphere of radius $M$. The solution of Eq.~(\ref{eq_dMi}) lies on the intersection of the two surfaces, as displayed in Fig.~\ref{fig_sphere}. If $M^2 \notin \left[2I_1E,2I_3E\right]$ then there is no solution. The lower-left panel of Fig.~\ref{fig_sphere} shows the evolution of $\vec{M}(t)=(M_1(t),M_2(t),M_3(t))$ given by Eq.~(\ref{eq_dMi}),  which is plotted on the sphere of radius $M$ for different energy levels. We can distinguish two different cases. The oscillating and the rotating solutions occur in the case  $2I_2E<M^2<2I_3E$ and $2I_1E<M^2<2I_2E$, respectively. The separatrix is given by $2I_2E=M^2$.

As mentioned above, the vector $\vec{M}$ can be expressed in terms of the two angles $\theta$ and $\psi$. The components of $\vec{M}$ in the rotating frame $(\vec{e}_1,\vec{e}_2,\vec{e}_3)$ are given by:
\begin{equation}
M_1 = M\cos\theta,\quad M_2=-M\sin\theta\cos\psi,\quad M_3=M\sin\theta\sin\psi. \label{eq_Mi}
\end{equation}
These equations are sufficient to show that, for a flip $\Delta\psi\simeq\pi$, the variation of $\theta$ is $\Delta\theta\simeq 0$. Indeed, we see on the lower panels of Fig.~\ref{fig_sphere} that $\psi_f\simeq\pi+\psi_0$ leads to $(M_{1f},M_{2f},M_{3f})\simeq(M_{10},-M_{20},-M_{30})$. Using Eq.~(\ref{eq_Mi}), we get $\theta_f\simeq\theta_0$ in this case.
A qualitative interpretation of the angles $\theta(t)$ and $\psi(t)$ can also be made by using the lower panels of Fig.~\ref{fig_sphere}. In the case of rotating solutions, it can be shown that the flip angle $\psi$ is not bounded, and $\theta$ belongs to $\in[0,\pi/2]$ or $[\pi/2,\pi]$. From Fig.~\ref{fig_angles}, we deduce that the handle along $\vec{e}_1$ stays above (resp. below) the plane $(X,Y)$, while the racket is rotating about its handle. Using the symmetries of the problem, we focus only on the case $\theta\in[0,\pi/2]$ for a rotating dynamics. For an oscillating solution, we have $\psi\in[0,\pi]$ or $\psi\in[-\pi,0]$, which means that the racket oscillates about its handle. Since $\theta\in[0,\pi]$, the handle of the racket can cross the plane $(X,Y)$. Figure \ref{fig_sphere} shows the three equilibrium points $p_+$, $p_-$ and $p_D$ of the dynamics. The point $p_+$ corresponds to the case where $\vec{M}$ is collinear to $\vec{e}_1$. Around this stable equilibrium point, the racket rotates only about its handle. The second stable rotation is associated with the point $p_-$ and a rotation of the racket about its $\vec{e}_3$- axis. Finally, the point $p_D$ corresponds to the unstable rotation about its $\vec{e}_2$- axis.
Note that the lower-left panel of Fig.~\ref{fig_sphere} depends on the shape of the rigid body. For a symmetric body for which $I_1=I_2=I_3$, the energy ellipsoid becomes a sphere. In the case $I_2=I_3$, the oscillating solution does not exist. The intersection of the two surfaces is a circle centered on $\vec{e}_1$, and the separatrix lies on the equator of the sphere. In this situation, the angle $\theta$ is constant and the flip angle $\psi$ evolves linearly with time~\cite{landau}. For $I_1=I_2$, there are only oscillating solutions and the separatrix is a circle in the $(\vec{e}_1,\vec{e}_2)$- plane. 
\section{The Full Euler equations\label{sect_euler_eq}}
The full Euler equations give the differential system satisfied by the three Euler angles $\theta$, $\phi$ and $\psi$ (see, e.g., the approach used in Ref.~\cite{landau}):
\begin{equation}
\begin{array}{lll}
& \dot{\theta}=M\left(\frac{1}{I_3}-\frac{1}{I_2} \right)\sin\theta\sin\psi\cos\psi \\
& \dot{\phi}=M\left(\frac{\sin^2\psi}{I_3}+\frac{\cos^2\psi}{I_2} \right) \\
& \dot{\psi}=M\left( \frac{1}{I_1}-\frac{\sin^2\psi}{I_3}-\frac{\cos^2\psi}{I_2}\right)\cos\theta.
\end{array}\label{eq_euler_angle}
\end{equation}
We introduce the following parameters to simplify the differential system:
\[a=\frac{I_2}{I_1}-1,\quad b=1-\frac{I_2}{I_3},\quad \gamma=\frac{2I_2E}{M^2}-1.\]
The parameters $a$, $b$ and $\gamma$ are easily interpretable. The coefficient $\gamma$ tells us how far we are from the separatrix, given by $\gamma=0$. The cases $\gamma<0$ and $\gamma>0$ correspond respectively to the oscillating and rotating behaviors. The coefficients $a$ and $b$ are associated with the symmetries of the rigid body. For instance, $a=0$ (or $b=0$) means that the solid is a symmetric top, and $a=b=0$ corresponds to a body with a spherical symmetry. The case where $a\gg 1$ and $b\gg 1$ gives the constraint $I_1\ll I_2\ll I_3$ on the moments of inertia.

The TRE is a geometric phenomenon which can be described independently of the time evolution of Euler angles. This permits to reduce the system under consideration to a planar system, a point which was not clearly identified in~\cite{MSA91}. The dynamical equations of the angles $\theta$ and $\psi$ can be expressed with respect to $\phi$ as follows:
\begin{equation}
\begin{array}{ll}
& \frac{d\theta}{d\phi} = -\frac{b\sin\theta\sin\psi\cos\psi}{1-b\sin^2\psi} \\
& \frac{d\psi}{d\phi} = \frac{(a+b\sin^2\psi)\cos\theta}{1-b\sin^2\psi}.
\end{array}
\label{eq_Euler}
\end{equation}
This system admits the following constant of motion, which can be derived by substituting~(\ref{eq_Mi}) into~(\ref{eq_first_int}):
\begin{equation}
\gamma = a-\sin^2\theta(a+b\sin^2\psi).\label{eq_gamma}
\end{equation}
The assumption $I_1<I_2<I_3$, and the fact that $2I_1E\leq M^2\leq 2I_3E$ lead to the relations:
\[0\leq b\leq 1,\quad 0\leq a\leq +\infty,\quad -b\leq \gamma\leq  a. \]
The two equilibrium points, given by $\gamma=a$ and $\gamma=-b$, correspond to the points $p_+$ and $p_-$ of Fig.~\ref{fig_sphere}, respectively.
Indeed, the parameter $\gamma$ takes the value $a$ for $\theta=0$ and $\gamma=-b$ occurs for $\theta=\psi=\pi/2$.

The two equations of the system~(\ref{eq_Euler}) govern respectively the motion of the handle and the twisting effect. After eliminating the $\cos\theta$ factor from Eq.~(\ref{eq_gamma}), only the second equation is thus necessary to describe the TRE.
\section{General description  of the evolution of the system}\label{sec4}
This paragraph is aimed at giving a general overview about the dynamics of the rigid body. We have seen qualitatively on Fig.~\ref{fig_sphere} that the oscillating or rotating properties of the system are related to the dynamics of the flip angle $\psi$, given by the second equation of~(\ref{eq_Euler}).

\begin{figure}[h!]
\centering
\includegraphics[scale=0.6]{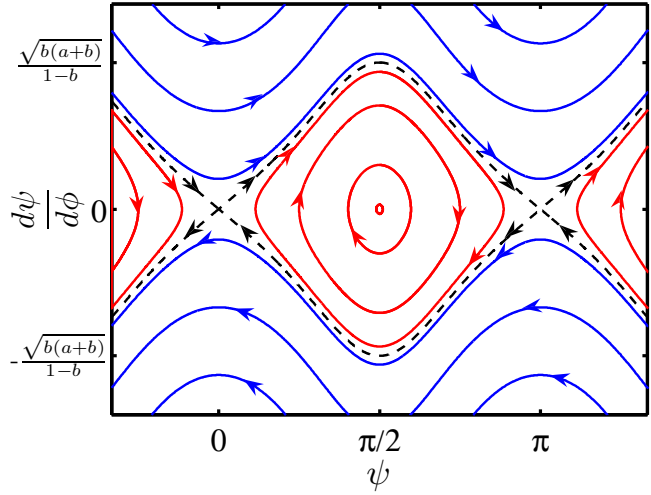}
\caption{(Color online) The flip phase portrait: The red (light gray) lines correspond to the oscillating solutions ($\gamma<0$), while the blue ones (dark gray) are associated with the rotating ones ($\gamma>0$). The separatrix ($\gamma=0$) is represented by the black dotted line. Two unstable equilibrium points belong to the separatrix. On these unstable points, the racket rotates exactly about its intermediate axis $\vec{e}_2$.\label{fig_ph_portrait}}
\end{figure}

Starting from Eq.~(\ref{eq_gamma}), we obtain:
\[\cos\theta=S\sqrt{\frac{\gamma+b\sin^2\psi}{a+b\sin^2\psi}},\]
where $S$ is the sign of $\cos\theta$. We deduce that the dynamics of the flip angle satisfies:
\begin{equation}
\frac{d\psi}{d\phi} = S\frac{\sqrt{a+b\sin^2\psi} \sqrt{\gamma+b\sin^2\psi}}{1-b\sin^2\psi}.\label{eq_dpsidphi}
\end{equation}
Figure~\ref{fig_ph_portrait} displays the phase portrait of the angle $\psi$.
Two equilibrium points can be identified: A stable one for $\psi=\pi/2(\mod\pi)$ and $d\psi/d\phi=0$, and an unstable one for $\psi=0(\mod\pi)$ and $d\psi/d\phi=0$.
These two points are associated with the stable rotation about $\vec{e}_3$ and the unstable one about $\vec{e}_2$, respectively. For $\gamma>0$, $d\psi/d\phi$ is always positive (or negative), so that $S$ is constant ($S$ is also the sign of $d\psi/d\phi$). For $\gamma<0$, the sign of $d\psi/d\phi$ changes when $\psi$ is extremal. These extremal values can be found by noting that if $\gamma<0$, the right term in the square root of Eq.~(\ref{eq_dpsidphi}) is positive only if $\sin^2\psi_{\min}=\sin^2\psi_{\max}=|\gamma|/b$. It is then easy to show that each oscillating trajectory is bounded by:
\begin{equation}
\psi_{\min}=\arcsin [\sqrt{\frac{|\gamma|}{b}}],\quad \psi_{\max}=\pi-\arcsin [\sqrt{\frac{|\gamma|}{b}}],\label{eq_borne_psi}
\end{equation}
and the sign of $d\psi/d\phi$ changes when $\psi$ takes one of these two values. The phase portrait of Fig.~\ref{fig_ph_portrait} is very similar to the standard phase portrait of a pendulum \cite{Golstein50}, except for the fact that the racket can oscillate about its handle with an amplitude which is at most of $\pi$, while this value is $2\pi$ for a pendulum. In Fig.~\ref{fig_ph_portrait}, it can be seen that the only way to connect two unstable equilibrium points is to make a variation of $\Delta\psi=\pi$. Roughly speaking, the flip effect can be viewed as a trajectory in a neighborhood of the separatrix, which goes from a point close to an unstable state to a point near another unstable state. \\

\begin{figure}[h!]
\centering
\includegraphics[scale=0.6]{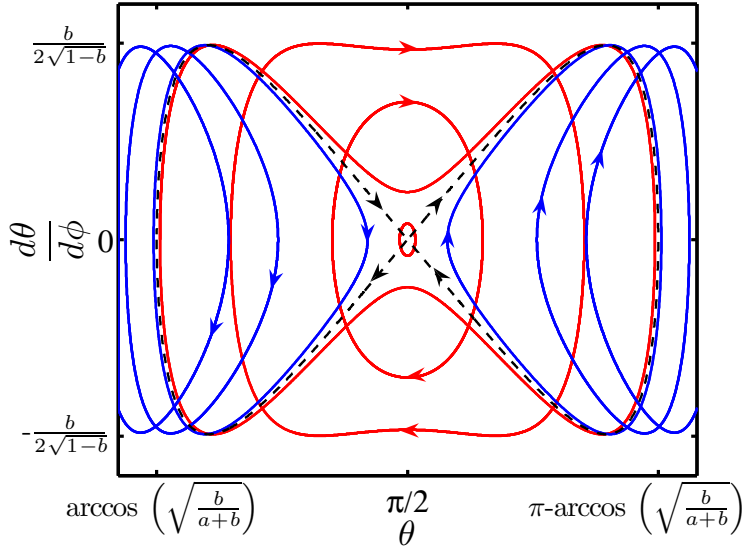}
\caption{(Color online) Phase portrait for the angle $\theta$. The red (light gray) lines correspond to the oscillating solutions ($\gamma<0$), the blue ones (dark gray) are associated with the rotating ones ($\gamma>0$), and the separatrix ($\gamma=0$) is represented by the black dotted line.\label{fig_portrait_theta}}
\end{figure}

\begin{figure}[h!]
\centering
\includegraphics[scale=0.6,trim=80 80 70 35,clip]{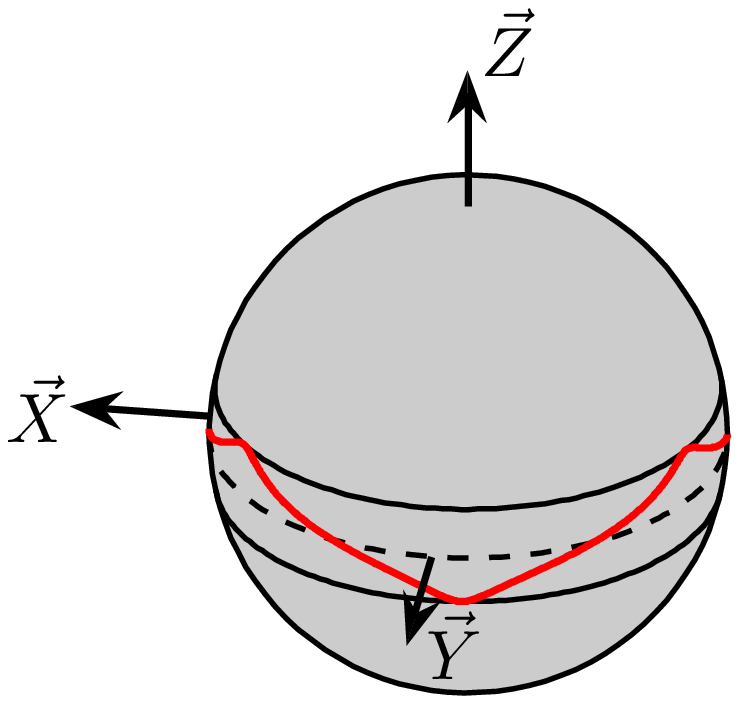}
\includegraphics[scale=0.6,trim=80 80 70 35,clip]{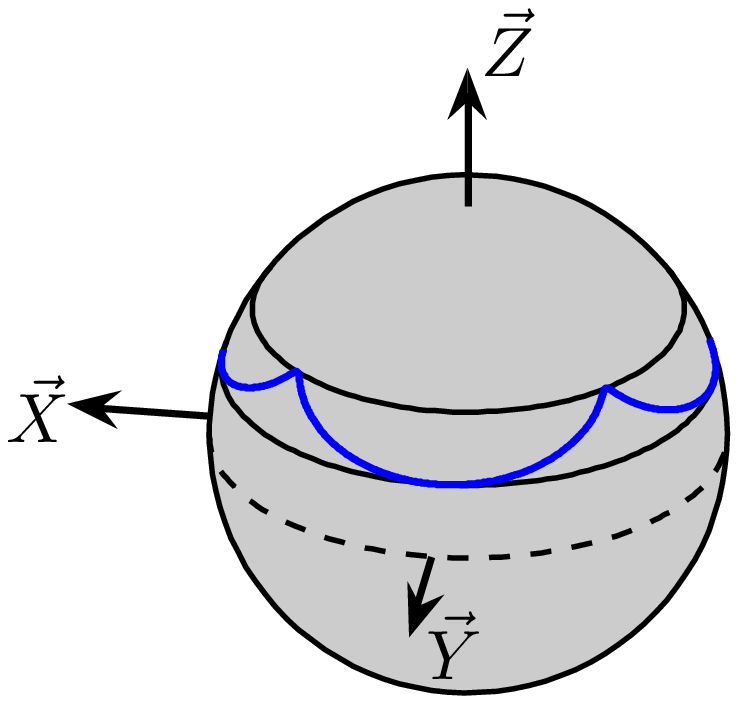}
\caption{(Color online) Illustration of the trajectory of the handle on the sphere for $\gamma<0$ (left panel) and $\gamma>0$ (right panel). The solid lines indicate the bounds of the angle $\theta$.\label{fig_thetamin}}
\end{figure}

We can make the same analysis for the first equation of~(\ref{eq_Euler}). Using Eq.~(\ref{eq_gamma}), we can show that:
\[\sin\psi\cos\psi=\frac{S_2}{b\sin^2\theta}\sqrt{(\gamma+b-(a+b)\cos^2\theta)(a\cos^2\theta-\gamma)},\]
with $S_2$ the sign of $\sin\psi\cos\psi$. We deduce that the dynamics of $\theta$ is governed by the following differential equation:
\begin{equation}
\frac{d\theta}{d\phi}=-S_2\sin\theta\frac{\sqrt{(\gamma+b-(a+b)\cos^2\theta)(a\cos^2\theta-\gamma)}}{\gamma+1-(a+1)\cos^2\theta}.
\label{eq_dthetadphi}
\end{equation}
The phase portrait $\theta(\phi)$ is displayed on Fig.~\ref{fig_portrait_theta}.
Note that we cannot distinguish on Fig.~\ref{fig_portrait_theta} the stable and the unstable points, since they both occur for $(\theta,d\theta/d\phi)=(\pi/2,0)$. Moreover, the oscillating curves can cross the line of equation $\theta=\pi/2$, but not the rotating ones (see Sec.~\ref{sec3} for details). The bounds of the angle $\theta$ can be determined by using the fact that the argument of the square root of Eq.~(\ref{eq_dthetadphi}) has to be positive. We recall that, for $\gamma>0$, we only consider the case $\theta \in [0,\frac{\pi}{2}]$ (see Sec.~\ref{sec2}). Then, it is straightforward to show that:
\begin{equation}
\begin{array}{ll}
\textrm{For }\gamma>0: & \quad \textrm{For }\gamma<0: \\
\theta_{\min}=\arccos [\sqrt{\frac{\gamma+b}{a+b}}] &\quad \theta_{\min}=\arccos [\sqrt{\frac{\gamma+b}{a+b}}]\\
\theta_{\max}=\arccos [\sqrt{\frac{\gamma}{a}}] &\quad \theta_{\max}=\pi-\arccos [\sqrt{\frac{\gamma+b}{a+b}}]
\end{array}
\label{eq_borne_theta}
\end{equation}
An illustration of the motion of the handle on the sphere is given on Fig.~\ref{fig_thetamin}.
\section{Analytical study of the Tennis racket effect \label{sect_gamma0}}
The TRE is a trajectory of the racket such that $\Delta\psi|_{\Delta\phi=2\pi}\simeq\pi$. We are therefore interested in the dynamics of $\psi$ with respect to $\phi$, given by Eq.~(\ref{eq_dpsidphi}).

In a TRE experiment, we start with initial conditions such that $\psi_0\simeq 0$ and $(d\psi/d\phi)_0\simeq 0$, that is in a neighbourhood of the unstable equilibrium point of the phase portrait of Fig.~\ref{fig_ph_portrait}. The system follows in this case a trajectory close to the separatrix. Since Eq.~(\ref{eq_dpsidphi}) cannot be integrated in terms of elementary  functions \cite{abramowitz}, we prefer to use an asymptotic expansion around $\gamma=0$ to analytically describe the behavior of the solution. We first explain the dynamics along the separatrix. If the initial point of the trajectory belongs to the separatrix then the system will follow this curve and will reach one of the unstable equilibrium points asympotically in an infinite time.

Equation~(\ref{eq_dpsidphi}) leads to:
\begin{equation}
\int_{\psi_0}^{\psi_f}f(a,b,\psi,\gamma)d\psi=\Delta\phi=2\pi,
\label{eq_robust_gen}
\end{equation}
with
\[f(a,b,\psi,\gamma)=S\frac{1-b\sin^2\psi}{\sqrt{a+b\sin^2\psi}\sqrt{\gamma+b\sin^2\psi}}.\]
In the following, we choose to focus on the symmetric solutions characterized by:
\begin{equation}
\psi_0=\epsilon,\; \psi_f=\pi-\epsilon,\label{eq_psi0psif}
\end{equation}
where $\epsilon$ is a small parameter. Note that this parameter can be interpreted as a defect of an ideal $\pi$- flip. This choice is motivated by the fact that, for a variation of $2\pi$ of the angle $\phi$, the maximum of variation of $\psi$ is given by the symmetric motion where $\psi_f=\pi-\psi_0$. The general analysis could be made along the same lines, but would be more cumbersome.

\begin{figure}[h!]
\centering
\includegraphics[scale=0.6]{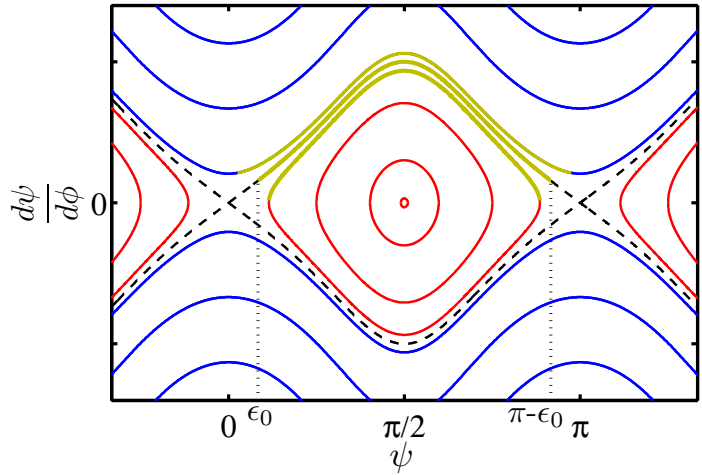}
\caption{(Color online) Illustration in brown (thick solid gray)of the symmetric curves described by Eq.~(\ref{eq_psi0psif}). Such lines correspond to a full rotation of the handle in the air, i.e. $\Delta\phi=2\pi$.\label{fig_symtraj}}
\end{figure}

In the symmetric situation, the variation of the twist is given by
\begin{equation}\label{eqdefect}
\left.\Delta\psi\right|_{\Delta\phi=2\pi}=\pi-2\epsilon,
\end{equation}
with $\epsilon=0$ corresponding to an exact $\pi$- flip. We are interested in trajectories near the separatrix where $\gamma =0$. For that reason, we express the parameter $\epsilon$ in the form of an asymptotic expansion in terms of $\gamma$:
\begin{equation}
\epsilon=\epsilon_0+\gamma\left.\frac{d\epsilon}{d\gamma}\right|_{\gamma=0}+O(\gamma^2).
\label{eq_epsilon_gen}
\end{equation}
Figure~\ref{fig_symtraj} shows the trajectories which are described by Eq.~(\ref{eq_epsilon_gen}). This formula will allow us to describe the robustness properties of the TRE with respect to the shape of the rigid body and the initial conditions of the dynamics. The different terms of the expansion have the following interpretation. The term $\epsilon_0$ depends on the shape of the body only, that is the parameters $a$ and $b$. The term $\gamma d\epsilon/d\gamma|_{\gamma=0}$ represents the variation of $\epsilon$ with respect to $\gamma$, i.e. its variation with respect to the initial conditions.

The asymptotic behavior of the flip is described by the following theorem:
\begin{theo}\label{prop1}
The flip defect $\epsilon$ given by Eq.~(\ref{eqdefect}) can be expressed as follows:
\begin{equation}
\epsilon(\gamma)=\epsilon_0-\frac{\gamma}{4b\epsilon_0}+O(\gamma^2),
\label{eq_pavao}
\end{equation}
where
$$
\epsilon_0=e^{-\sqrt{ab}\pi}\left(2\sqrt{\frac{a}{a+b}}e^{-\frac{b}{\sqrt{1+b/a}}}+o\left(e^{-\sqrt{ab}\pi}\right)\right).
$$

\end{theo}
\begin{rem} The flip defect, $\epsilon_0$, on the separatrix  tends exponentially to zero, for $ab\to\infty$. For fixed parameters $a,~b$, the flip defect goes to $\epsilon_0$, as $\gamma$ goes to zero, with a speed given by (\ref{eq_pavao}).
\end{rem}
The proof of Theorem~\ref{prop1} is detailed in the~\ref{app}. Note that, without any approximation, the parameter $\epsilon_0$ is obtained by solving the following equation (see~\ref{app}):
\begin{equation}
\frac{\sqrt{1+\frac{b}{a}\sin^2\epsilon_0}-\cos\epsilon_0}{\sqrt{1+\frac{b}{a}\sin^2\epsilon_0}+\cos\epsilon_0}= e^{-2\sqrt{ab}\left[\pi+\arcsin\left(\sqrt{\frac{b}{a+b}}\cos\epsilon_0\right)\right]}.
\end{equation}
Thus, for a symmetric top such that $b\to 0$, we get that $\epsilon_0\to \pi/2$, which means that there is no flip along the separatrix since $\Delta\psi|_{\Delta\phi=2\pi}=\pi-2\epsilon_0=0$. This behavior is not surprising because it can be shown on the phase portrait of Fig.~\ref{fig_ph_portrait} that, in this limiting case,   the separatrix is a horizontal line of equation $d\psi/d\phi=0$. We deduce that each point of this separatrix is an equilibrium point and for $b\to 0$, the only way to have a symmetric curve such that $\psi_0=\epsilon_0$ and $\psi_f=\pi-\epsilon_0$ is to choose $\epsilon_0=\pi/2$.

\begin{figure}[h]
\centering
\includegraphics[scale=0.6]{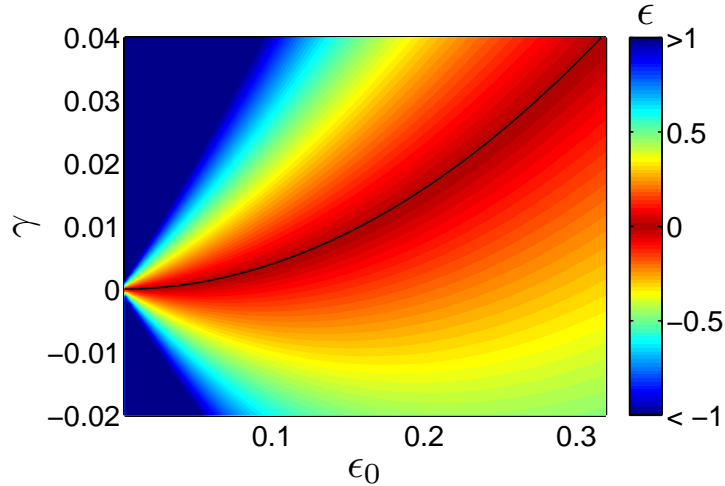}
\caption{Evolution of $\epsilon(\gamma)$ as a function of $\epsilon_0$ and $\gamma$ for $b=0.1$. The black line corresponds to an exact TRE ($\epsilon=0$).\label{fig_roban}}
\end{figure}

This asymptotic analysis can also be used to derive an approximate formula to estimate the flip effect. This formula shows that the larger the product $ab$ is, the smaller the parameter $\epsilon_0$ is. For a standard tennis racket \cite{standard}, we have:
\[\epsilon_{0}=0.1150\textrm{ rad},\]
i.e. a flip of $\Delta\psi_0=\pi-2\epsilon_{0}=\pi(1-0.073)$. Without any approximation, we obtain the same value to an accuracy of $10^{-3}$.
In contrast, the larger the term $ab$ is, the less robust is the effect. Indeed, the second term of Eq.~(\ref{eq_pavao}) is very sensitive to $\gamma$ if $\epsilon_0$ is small. In the case of Ref.~\cite{standard}, we have:
\[\left.\frac{d\epsilon}{d\gamma}\right|_{\gamma=0}=-34.5.\]
If we assume that the TRE is satisfied if
\begin{equation}
\Delta\psi|_{\Delta\phi=2\pi}\in[\pi-\pi/6,\pi+\pi/6],
\label{eq_TRE}
\end{equation}
then, $\epsilon$ being related to $\Delta\psi$ via $\Delta\psi=\pi-2\epsilon$, we get that $\epsilon\in[-\pi/12,\pi/12]$. For the standard tennis racket~\cite{standard}, we arrive at:
\[\gamma|_{\epsilon=-\pi/12}=1.09\times 10^{-2}, \gamma|_{\epsilon=\pi/12}=-4.2\times 10^{-3}.\]
Finally, in order to illustrate the formula~(\ref{eq_pavao}), we fix the parameter $b$ to $0.1$ and we represent the value of $\epsilon$ in the plane $(\epsilon_0,\gamma)$. Note that each value of $\epsilon_0$ is associated with one value of $a$ through the equation of $\epsilon_0$ in Theorem~\ref{prop1}. The result is displayed on Fig.~\ref{fig_roban}.
This figure shows that for small values of $\epsilon_0$, a TRE occurs close to the separatrix. We observe that a larger range of $\gamma$ satisfies the TRE for high values of $\epsilon_0$, which means that the effect is more robust in this case.

\section{Numerical description of the tennis racket effect} \label{sec5}
\begin{figure}[h!]
\centering
\includegraphics[scale=0.35,trim=0  48 30 15,clip]{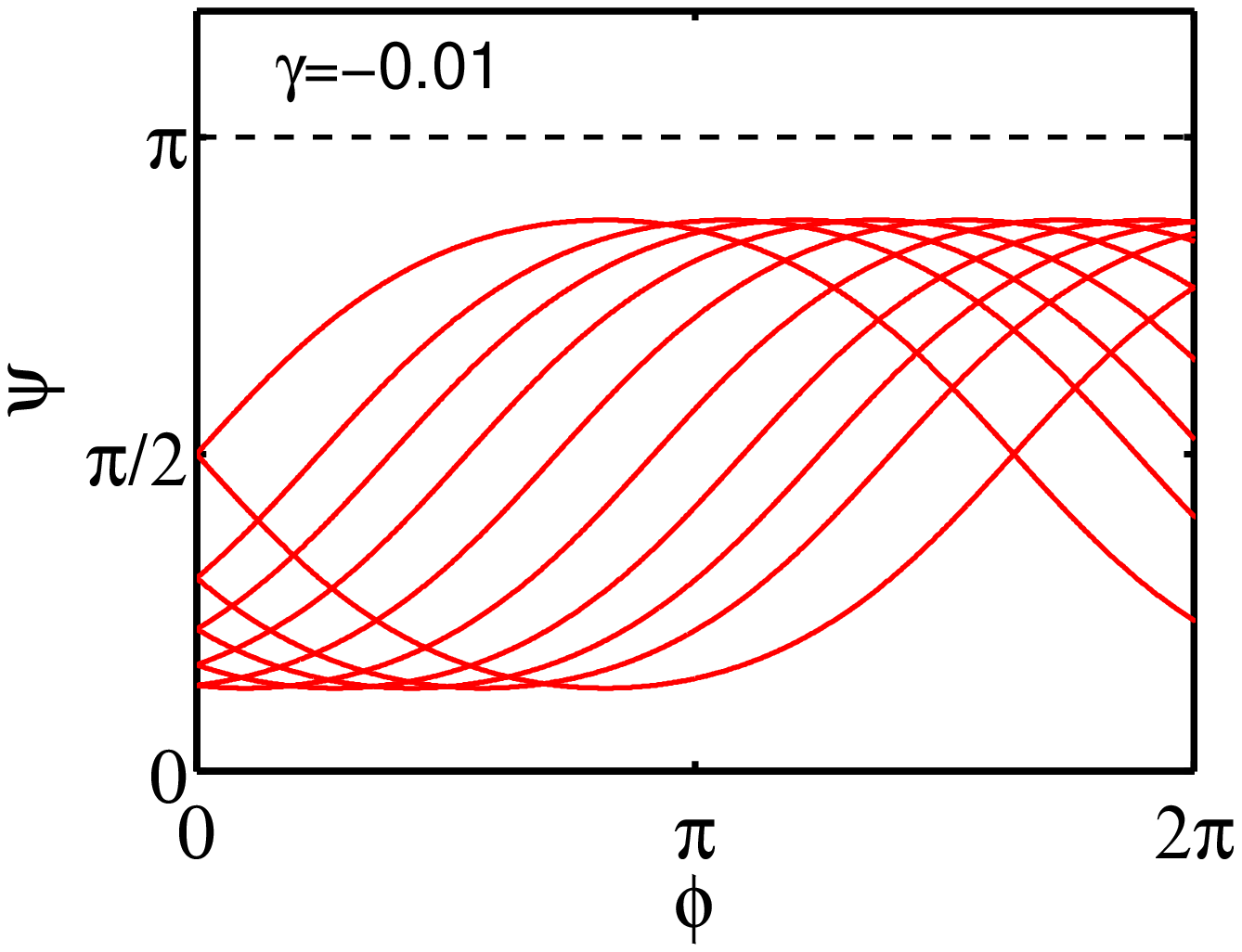}
\includegraphics[scale=0.35,trim=70 48 30 15,clip]{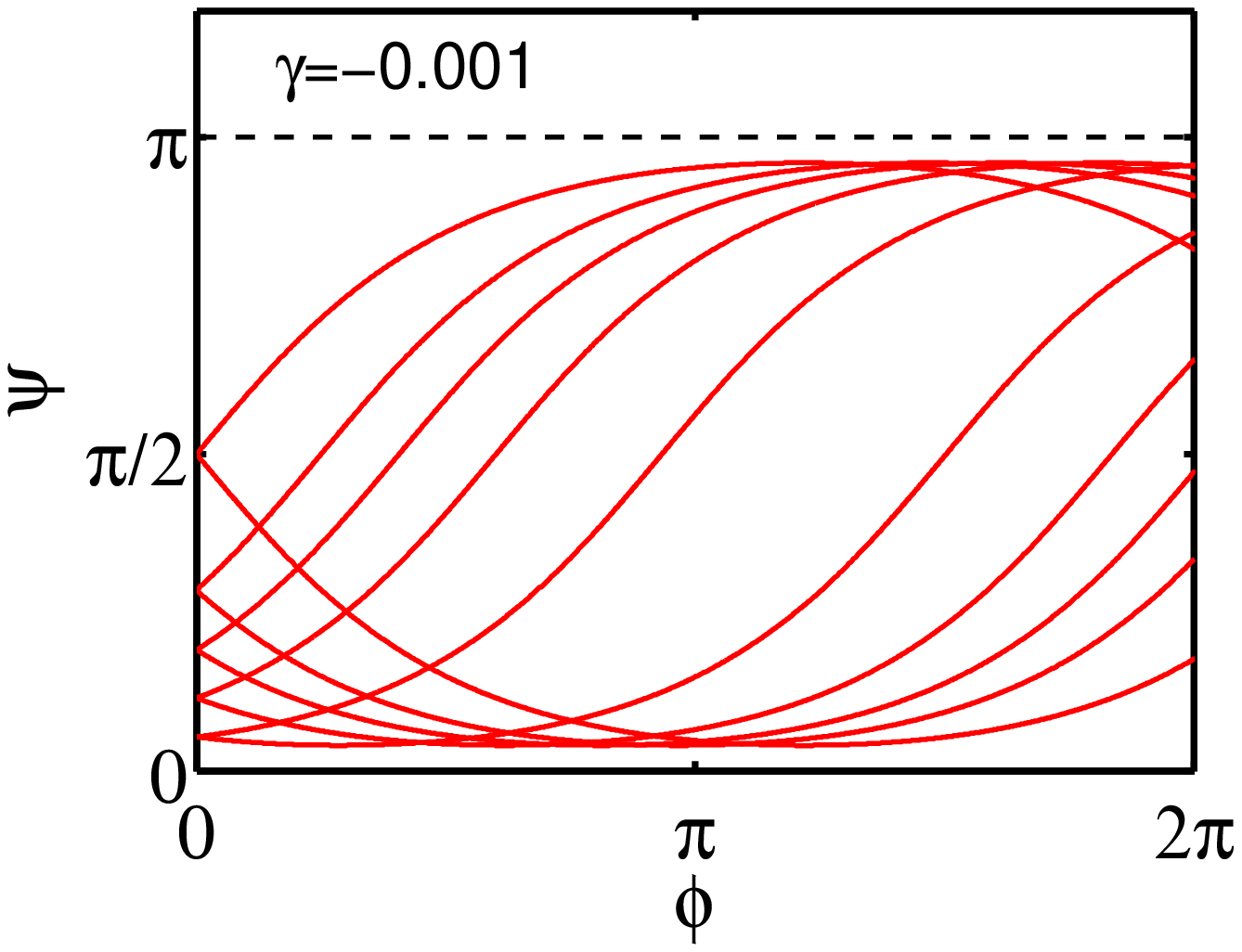}
\includegraphics[scale=0.35,trim=0  48 30 15,clip]{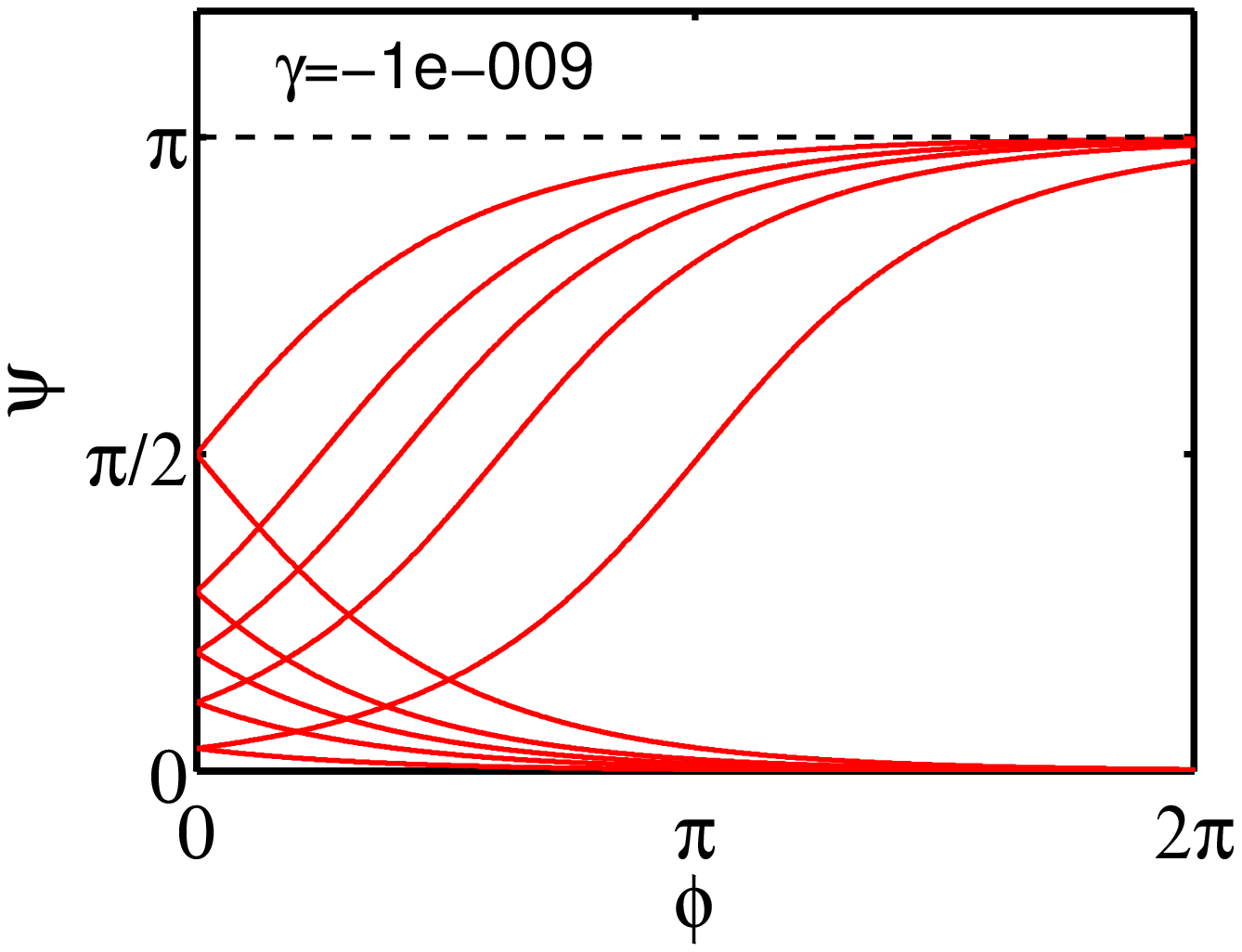}
\includegraphics[scale=0.35,trim=70 48 30 15,clip]{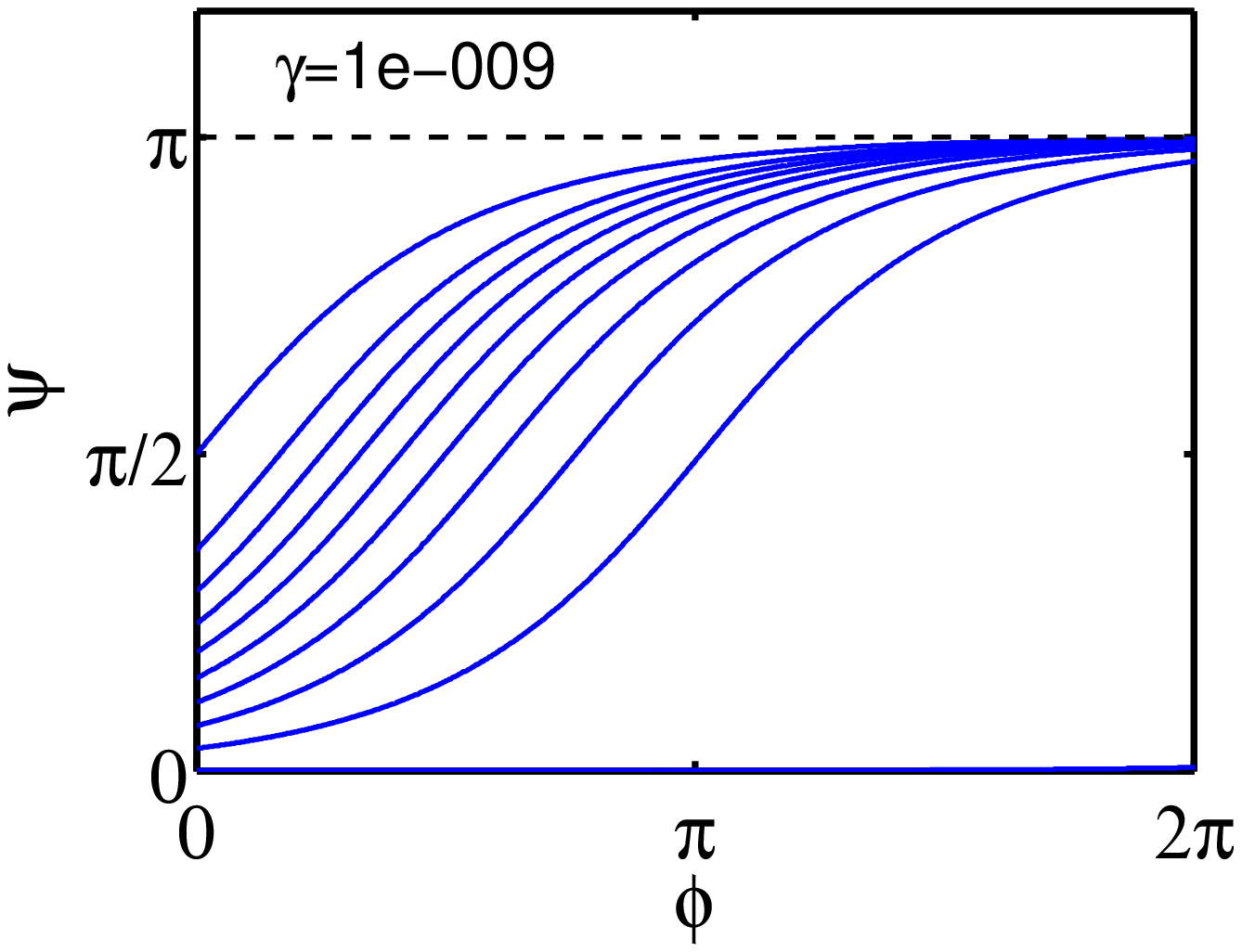}
\includegraphics[scale=0.35,trim=0  0  30 15,clip]{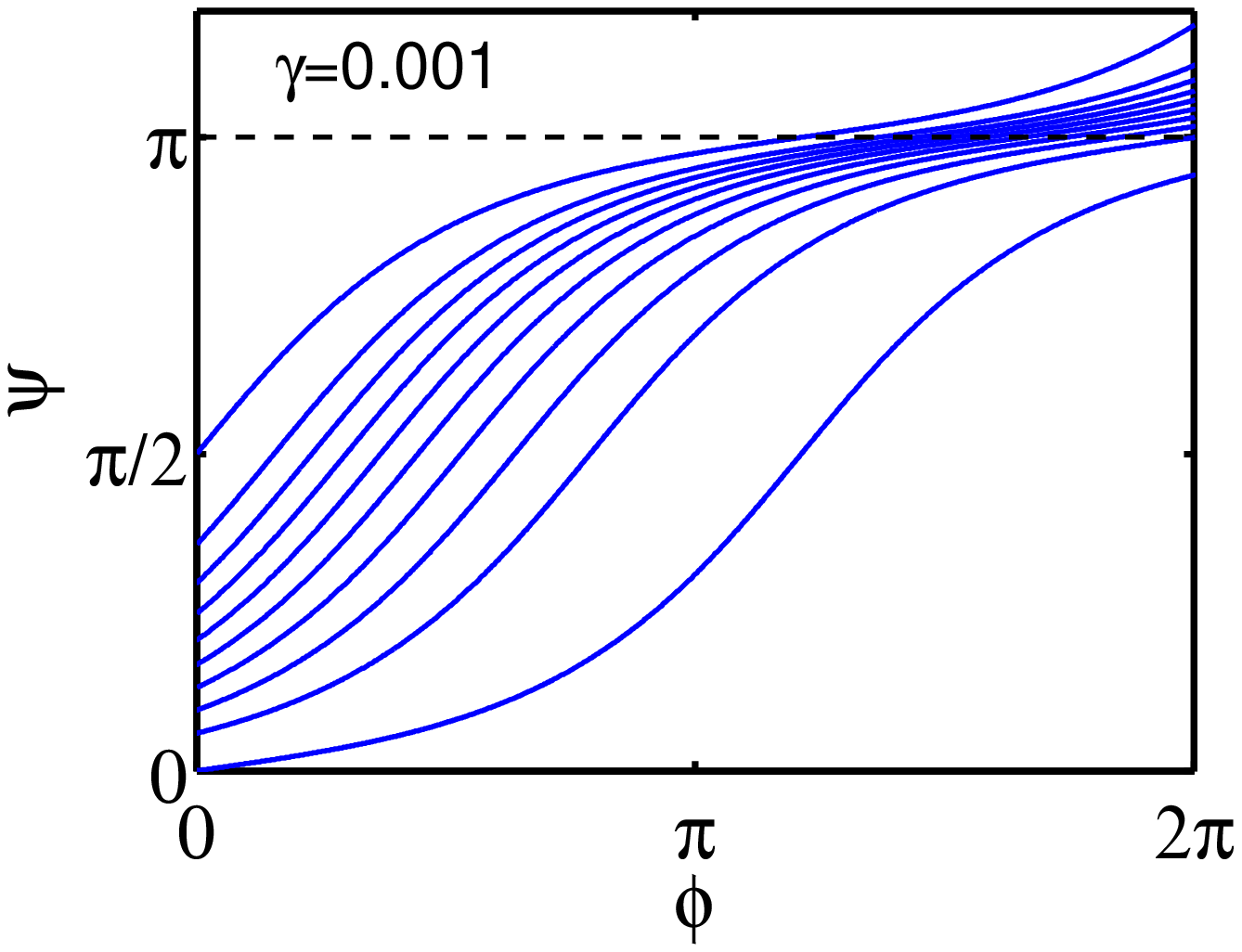}
\includegraphics[scale=0.35,trim=70 0  30 15,clip]{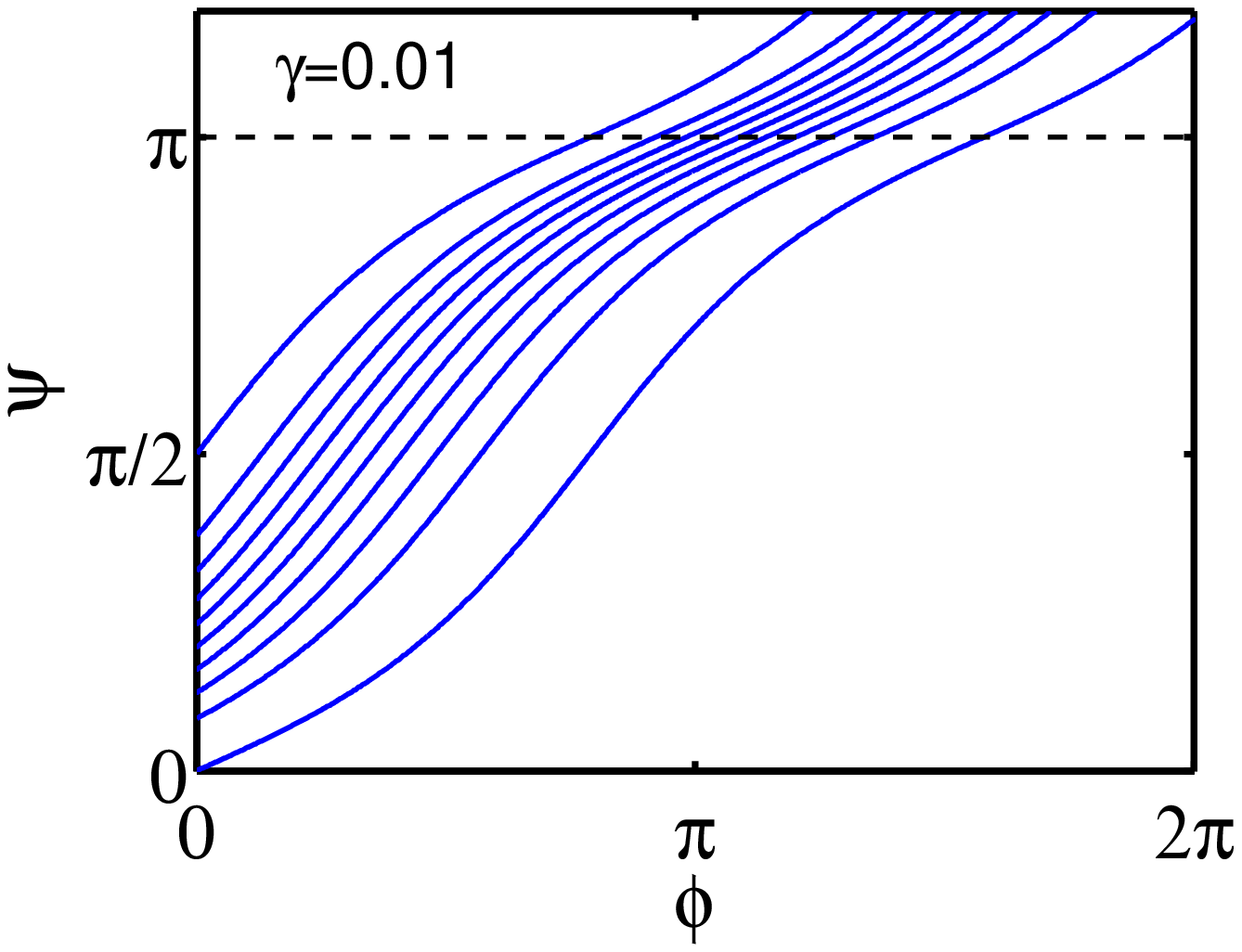}
\caption{(Color online) Plot of the trajectories $\psi=\psi(\phi)$ for different initial conditions and different values of $\gamma$. Red (light gray) and blue (dark gray) curves correspond to $\gamma<0$ and $\gamma>0$, respectively.\label{fig_psiphi}}
\end{figure}

\begin{figure}[h!]
\includegraphics[scale=0.5,trim=0 0 15  0,clip]{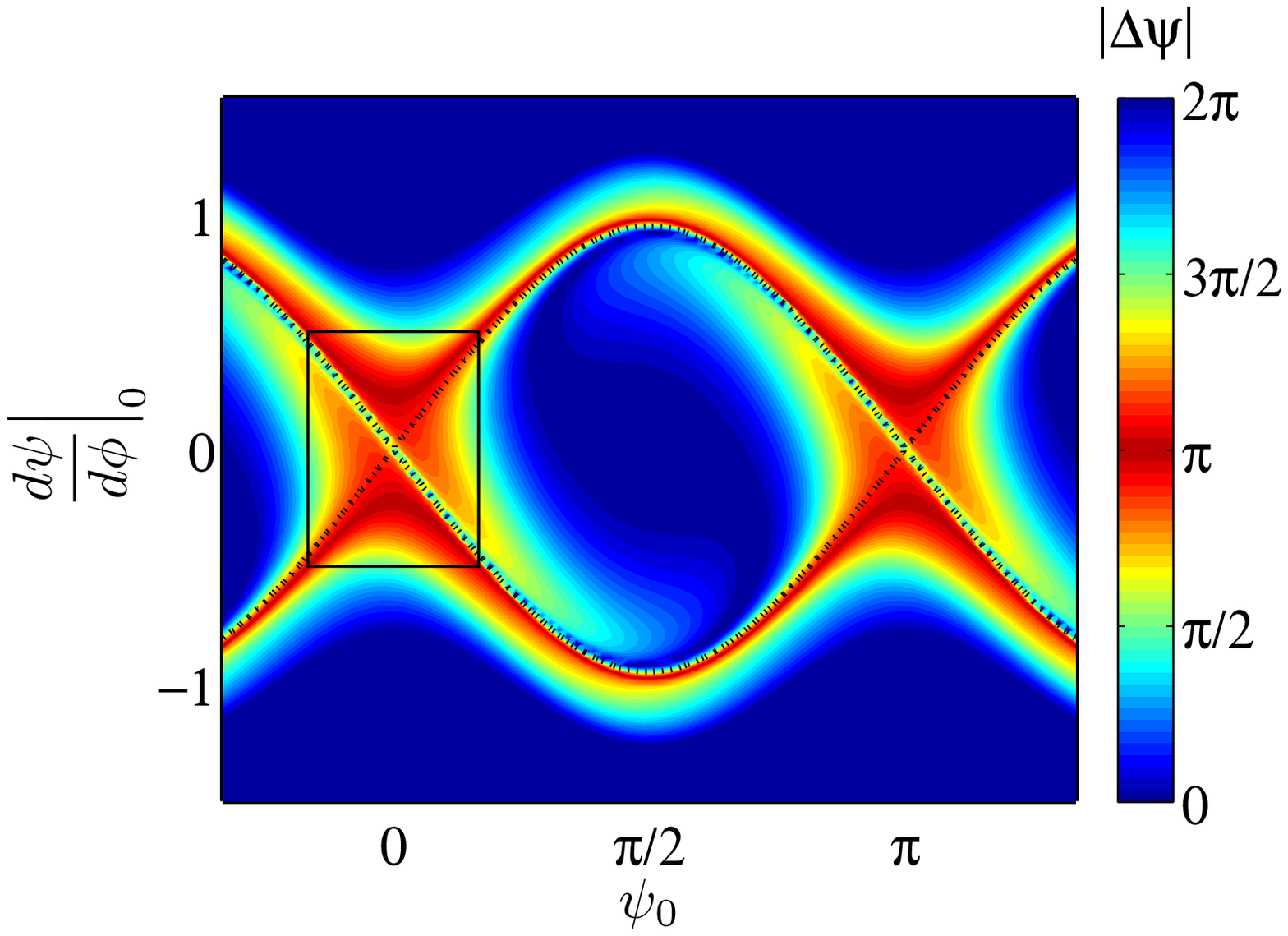}
\includegraphics[scale=0.4,trim=19 -70 0  0,clip]{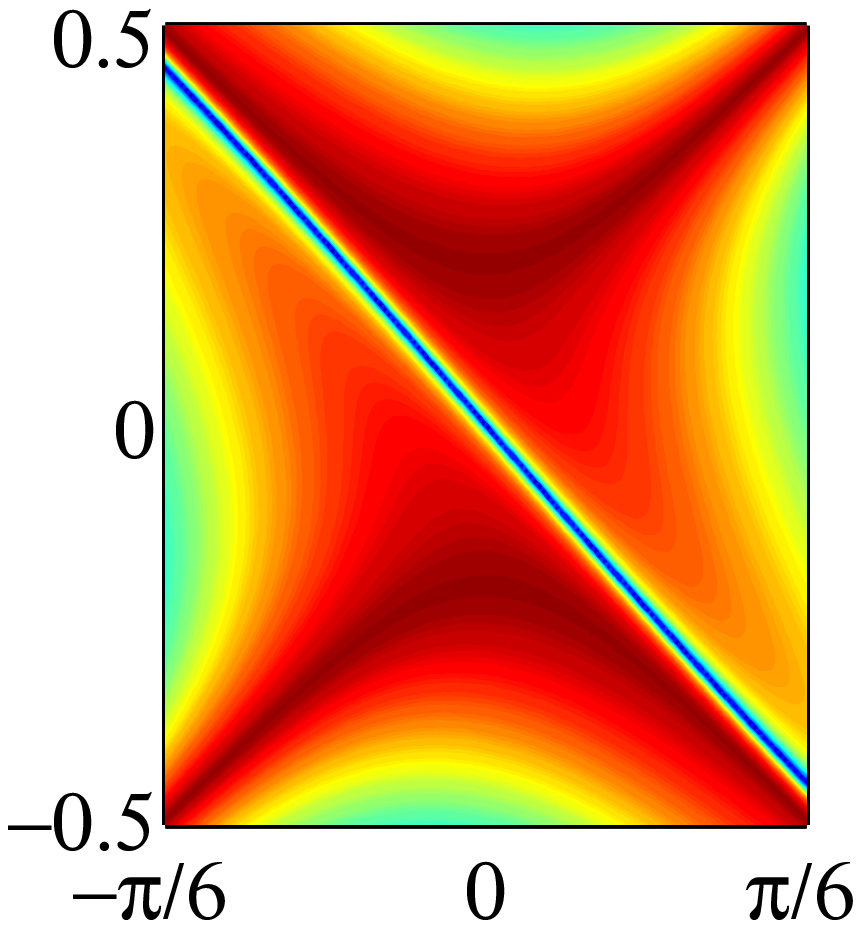}
\caption{(Color online) \textit{Left Panel:} Contour plot of the twist $|\Delta\psi|$ after a full rotation of the handle $\Delta\phi=2\pi$, for different initial conditions $(\psi_0,(d\psi/d\phi)_0)$. The black dotted line depicts the position of the separatrix for which $\gamma=0$. The solid rectangle represents a domain of initial conditions close to the unstable point. \textit{Right panel:} Zoom of the region around the point $(0,0)$ of the left panel. The separatrix has been removed for clarity. The color code is the same for the two panels. \label{fig_rob}}
\end{figure}

This paragraph focuses on a numerical study of the TRE. We consider the case of a standard tennis racket \cite{standard}. Setting a value of the parameter $\gamma$, we compute the trajectories of the system in the $(\phi,\psi)$- space for different values of $\psi_0$. The different trajectories are represented in Fig.~\ref{fig_psiphi}. We observe in Fig.~\ref{fig_psiphi} that the TRE does not occur far from the separatrix, approximately when $|\gamma|>10^{-2}$. In contrast, close to the separatrix, the dynamics exhibits an approximate flip of $\pi$ for a large range of initial values of $\psi_0$.
Figure~\ref{fig_rob} is a contour plot displaying the flip $|\Delta\psi|$ as a function of the initial conditions $(\psi_0,(d\psi/d\phi)_0)$. This set of initial conditions is the more intuitive one from an experimental point of view since $(d\psi/d\phi)_0$ is the initial angular velocity of the racket along the handle and $\psi_0$ the initial flip angle. The unstable rotation along the $e_2$- axis is characterized by $\psi_0=0\mod\pi$ and $(d\psi/d\phi)_0=0$. When we consider initial conditions close to this point, Fig.~\ref{fig_rob} shows that a near flip is very likely to happen. We see that the value of the flip obeys to a central symmetry with respect to the unstable points of the dynamics.
The right panel of Fig.~\ref{fig_rob} shows that the variation of the flip angle is approximately zero on the separatrix. This point is due to the fact that the time evolution of $\psi$ on this line is very slow and the TRE is defined for a time interval such that $\Delta\phi=2\pi$.

\begin{figure}
\centering
\includegraphics[scale=0.55]{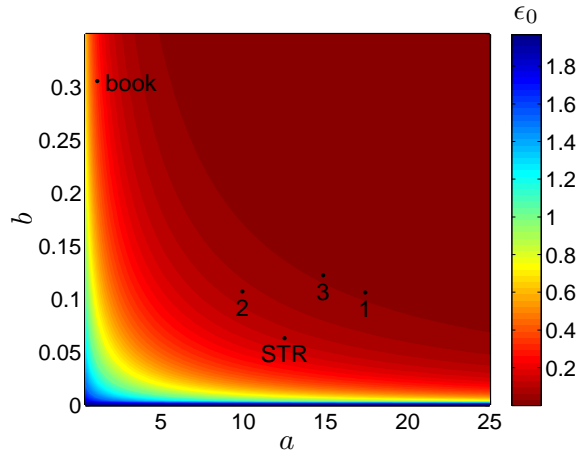}
\caption{(Color online) Plot of the parameter $\epsilon_0$ as a function of $a$ and $b$. The black dots correspond to the different objects of Tab.~\ref{tab1}.\label{fig_compar_racket}}
\end{figure}

For any rigid body, a TRE occurs along the separatrix if the parameter $\epsilon_0$ is small enough. Figure~\ref{fig_compar_racket} represents the value of $\epsilon_0$ with respect to the parameters $a$ and $b$. Note the hyperbolic shape of the $\epsilon_0$- level curves. We add on this figure different points corresponding to the standard tennis racket of~\cite{standard} and three other tennis rackets for which the principal moments of inertia are given in~\cite{nesbit}. We have also estimated the TRE for the famous book by Goldstein of Ref.~\cite{Golstein50}. The moments of inertia can be easily computed if we consider the book as a rectangular box. The three principal axes of inertia are along the length of the book ($\vec{e}_1$), along the width ($\vec{e}_2$) and perpendicular to the face of the book ($\vec{e}_3$). We find:
\[a=1.11,\; b=0.31.\]
The values of the flip angle $\Delta\psi|_{\gamma=0}=\pi-2\epsilon_0$ are given in Tab.~\ref{tab1}.
\begin{table}[t]
\caption{Value of the flip angle for different objects. The moments of inertia are expressed in kg.mm$^{-1}$.s$^{-2}$ and the angles in radian. The
characteristics of the STR are given in \cite{standard} and in \cite{nesbit} for the rackets 1, 2 and 3. The book is the Goldstein book \cite{Golstein50}.\label{tab1}}
\begin{tabular}{|cccccccc|}
\hline
Object & $I_1$ & $I_2$ & $I_3$ & $a$ & $b$ & $\epsilon_0$ & $\Delta\psi|_{\gamma=0}$\\
\hline
\hline
Racket (STR) & 1210 & 16380 & 17480 & 12.54 & 0.06 & 0.1150&$\pi(1-0.0732)$\\
Racket 1 & 886 & 16336 & 18270 & 17.44 & 0.11 & 0.0251&$\pi(1-0.0160)$\\
Racket 2 &1319 & 14445 & 16175 & 9.95 & 0.11 & 0.0699& $\pi(1-0.0445)$\\
Racket 3 &448 & 7126 & 1818 & 14.88 & 0.12 & 0.0255 &$\pi(1-0.0162)$\\
book & &  &  & 1.11 & 0.31 & 0.2139& $\pi(1-0.1362)$\\
\hline
\end{tabular}
\end{table}

\section{Conclusion and perspectives}\label{sec6}
The tennis racket effect was initially mathematically analyzed in \cite{MSA91}, and reproduced in \cite{cushman}. In this paper, we expand considerably this initial treatment by deriving exact and asymptotic formulae describing the flip of the head of the racket. We have shown that any rigid body (with some constraints on its moments of inertia) can exhibit this effect. We have analyzed in detail the robustness properties of the flip with respect to the shape of the solid and the chosen initial conditions. Extensive numerical computations have finally confirmed the geometric description made in the first part of this work.

More importantly, this paper demonstrates that this effect can occur in a variety of classical systems, whose dynamics is governed by Euler equations. This study opens the road to future work in several directions. An open problem is the extension of the present paper to quantum systems. A possible candidate to exhibit this geometric property at the quantum level could be asymmetric top molecules \cite{landauQ} in gas phase whose motion can be controlled to some extent by femtosecond laser pulses \cite{rouzee,lee}. An example is given by the Ethylene molecule for which rotational constants are: $B_1=4.64$ cm$^{-1}$, $B_2=1.001$ cm$^{-1}$, $B_3=0.828$ cm$^{-1}$. These constants are inversely proportional to the principal moments of inertia of the molecules. In the case $\gamma=0$, this leads to a classical flip of $\Delta\psi=\pi(1-0.0528)$. Another option could be to consider the dynamics of a linear chain of three coupled spins subjected to radio-frequency magnetic fields \cite{levitt,cat,lapert}. Previous studies have shown that the optimal controlled trajectories of this system are given by solutions of Euler equations \cite{vandamme,nimbalkar}. Work is in progress on these different points. A final aspect is the possible role of the TRE on a tennis player \cite{brody1,brody2}. In other words, one can ask whether this effect has to be maximized or minimized in order to improve the performance of the player \cite{nesbit,rogowski}, as well as to prevent injuries.

\appendix
\section{Proof of the formula describing the TRE}\label{app}
This appendix focuses on the mathematical derivation of the analytical formula of Theorem~\ref{prop1}. We recall that the idea is to express the angle $\epsilon$ close to $\gamma=0$ as the sum of two terms:
\[\epsilon(\gamma)=\epsilon_0+\gamma\left.\frac{d\epsilon}{d\gamma}\right|_{\gamma=0}.\]
The two terms $\epsilon_0$ and $d\epsilon/d\gamma|_{\gamma=0}$ are computed separately.
\subsection*{Computation of $\epsilon_0$}
We show in this paragraph how to compute the parameter $\epsilon_0$. We substitute $\gamma$ by 0 and we use the assumption~(\ref{eq_psi0psif}) in Eq.~(\ref{eq_robust_gen}). We obtain:
\begin{equation}
\int_{\epsilon_0}^{\pi-\epsilon_0}f(a,b,\psi,0)d\psi=2\pi,
\end{equation}
where
\[f(a,b,\psi,\gamma)=S\frac{1-b\sin^2\psi}{\sqrt{a+b\sin^2\psi}\sqrt{\gamma+b\sin^2\psi}}.\]
We start from a point of the separatrix such that $(d\psi/d\phi)_0>0$. Since the trajectory cannot cross the unstable equilibrium points (see Fig.~\ref{fig_ph_portrait}), $d\psi/d\phi$ does not change sign, and we have $S=+1$. Introducing the auxiliary variable:
\[x=\sqrt{\frac{b}{a+b}}\cos\psi, \]
we arrive, after some manipulations, at:
\[\int_{x_0}^{-x_0}\frac{dx}{(1-\frac{a+b}{b}x^2)\sqrt{1-x^2}}-b\frac{dx}{\sqrt{1-x^2}}=-2\pi b.\]
The integration gives:
\[\arctanh\left(\sqrt{\frac{a}{b}}\frac{x_0}{\sqrt{1-x_0^2}} \right)-\sqrt{ab}\;\arcsin(x_0)=\sqrt{ab}\;\pi.\]
The $\arctanh$ function can be expressed in terms of a logarithm with the formula $\arctanh(u)=\frac{1}{2}\ln[(1+u)/(1-u)]$. Taking the exponential and the inverse, we get:
\[\frac{\sqrt{1+\frac{b}{a}\sin^2\epsilon_0}-\cos\epsilon_0}{\sqrt{1+\frac{b}{a}\sin^2\epsilon_0}+\cos\epsilon_0}= e^{-2\sqrt{ab}\left[\pi+\arcsin\left(\sqrt{\frac{b}{a+b}}\cos\epsilon_0\right)\right]}.\]
$\epsilon_0=0$ is an asymptotic solution of this equation when $ab\rightarrow+\infty$. When $ab\gg 1$, the dominant term of this equation is the exponential term. Making a series expansion in $\epsilon_0$ on the left-hand side, we have:
\[\frac{1}{4}\left(1+\frac{b}{a}\right)\epsilon_0^2+hot=e^{-2\sqrt{ab}\left[\pi+\arcsin\left(\sqrt{\frac{b}{a+b}}\cos\epsilon_0\right)\right]}.\]
Taking the square root, we get:
\begin{equation}\label{eqnew}
\epsilon_0+hot=2\sqrt{\frac{a}{a+b}}e^{-\sqrt{ab}\left[\pi+\arcsin\left(\sqrt{\frac{b}{a+b}}\cos\epsilon_0\right)\right]}.
\end{equation}

We are interested in the leading term in the case the parameter $ab$ goes to infinity. We introduce the following new parameters:
$$
\lambda=\frac{1}{\sqrt{ab}}, \quad \mu=\sqrt{\frac{b}{a+b}}=\frac{\lambda b}{\sqrt{1+\lambda^2b^2}}.
$$
Note that $ab\rightarrow \infty$ corresponds to $\lambda\rightarrow 0$ and gives $\mu\rightarrow 0$.
Equation (\ref{eqnew}) reads:
\begin{equation}\label{equation}
\epsilon_0=2\sqrt{1-\mu^2} e^{-\frac{1}{\lambda}\left[\pi+\arcsin\left(\mu\cos\epsilon_0\right)\right]}.
\end{equation}
Recall that the function $f$ given by $f(s)=e^{-1/s}$, for $s\ne0$ and $f(0)=0$ is of class $C^\infty$, infinitely flat at the origin.
We search for a solution $\epsilon_0$ of the above equation in the form
\begin{equation}\label{u}
\epsilon_0=2\sqrt{1-\mu^2} e^{-\frac{1}{\lambda}\left[\pi+\arcsin\left(\mu\right)+u\right]}=2\sqrt{1-\mu^2} f\left(\frac{\lambda}{\pi+\arcsin\left(\mu\right)+u}\right),
\end{equation}
with $u$ a differentiable function of $\lambda,~\mu$, with $u(0,0)=0$. Note that $\epsilon_0(u,\lambda,\mu)$ is smooth at $(u,\lambda,\mu)=(0,0,0)$. Substituting (\ref{u}) into (\ref{equation}) and simplifying, we obtain:
\begin{equation}\label{equ}
G(u,\lambda,\mu):=u+\arcsin\mu-\arcsin(\mu\cos(\epsilon_0(u,\lambda\mu)))+hot=0.
\end{equation}
Note that $G$ is a $C^\infty$ function at $(u,\lambda,\mu)=(0,0,0)$, $G(0,0,0)=0$ and $\frac{\partial G}{\partial u}(0,0,0)\ne0$. Hence the implicit function theorem applies and gives the existence of a $C^\infty$
solution $u=u(\lambda,\mu)$ of (\ref{equ}), verifying $u(0,0)=0$.
This justifies the existence of a solution of (\ref{equation}) in the form (\ref{u}).
Note that $\lim_{\lambda\to0}(u,\mu)=(0,0)$ and hence $\lim_{\lambda\to0}\epsilon_0=0$. For any $\delta>0$, we get $\epsilon_0=o(e^{-(\pi-\delta)/\lambda})$, when $\lambda\to0$.
The last step consists in determining the dependence of $u$ with respect to $\lambda$. We use again the implicit function theorem. This leads to the relation:
$$
\frac{du}{d\lambda}=-\frac{\frac{dG}{d\lambda}}{\frac{dG}{du}}(0,0,0).
$$
We have $\frac{dG}{du}(0,0,0)=1$. On the other hand by the chain rule, we get:
$$
dG/d\lambda=\frac{\partial G}{\partial\lambda}+\frac{\partial G}{\partial\mu}\frac{\partial\mu}{\partial\lambda}.
$$
Since $\frac{\partial G}{\partial \lambda}(0,0,0)=0$, we deduce that $\frac{\partial\mu}{\partial\lambda}=0$. It follows that:
\begin{equation}\label{dudlam}
\frac{du}{d\lambda}(0,0,0)=0
\end{equation}
i.e. $u=O(\lambda^2)$. Substituting the expressions for $u$ and $\mu$ into (\ref{u}), we obtain:
\begin{equation}\label{eps0}
\epsilon_0=e^{-\sqrt{ab}\pi}\left(2\sqrt{\frac{a}{a+b}}e^{-\frac{b}{\sqrt{1+b/a}}}+o\left(e^{-\sqrt{ab}\pi}\right)\right).
\end{equation}


\subsection*{Computation of $d\epsilon/d\gamma|_{\gamma=0}$}
Using the symmetries of the angle $\psi$, we have:
\[\int_{\epsilon}^{\pi-\epsilon}f(a,b,\psi,\gamma)d\psi=2\int_{\epsilon}^{\frac{\pi}{2}}f(a,b,\psi,\gamma)d\psi.\]
The motion of $\psi$ with respect to $\phi$ can be written as follows:
\[F(a,b,\epsilon,\gamma)=\int_{\epsilon}^{\frac{\pi}{2}}f(a,b,\psi,\gamma)d\psi=\pi.\]
We have:
\begin{equation}
\left.\frac{\partial\epsilon}{\partial\gamma}\right|_{\gamma=0}=-\left.\left(\frac{\partial\epsilon}{\partial F}\frac{\partial F}{\partial \gamma}\right)\right|_{\gamma=0}.
\label{eq_depsdgamma_form}
\end{equation}
The term $\partial \epsilon/\partial F$ can be easily determined. Differentiating with respect to one of the bounds of the integral gives:
\[
\frac{\partial F}{\partial \epsilon}=-f(a,b,\epsilon,\gamma)\Rightarrow \left.\frac{\partial \epsilon}{\partial F}\right|_{\gamma=0}=-\frac{1}{f(a,b,\epsilon_0,0)}=-S\sqrt{b}\sin\epsilon_0\frac{\sqrt{a+b\sin^2\epsilon_0}}{1-b\sin^2\epsilon_0}.
\]
With the assumption $\epsilon_0\ll 1$, the derivative can be expressed as follows:
\begin{equation}
\left.\frac{\partial \epsilon}{\partial F}\right|_{\gamma=0}=-S\sqrt{ab}\;\epsilon_0 + O(\epsilon_0^2).
\end{equation}
The term $\partial F/\partial \gamma$ is more difficult to compute. The uniform convergence of the integral $F$ gives:
\[\left.\frac{\partial F}{\partial \gamma}\right|_{\gamma=0}=\int_{\epsilon_0}^{\frac{\pi}{2}}\left.\frac{\partial}{\partial\gamma}f(a,b,\psi,\gamma)\right|_{\gamma=0}d\psi.\]
The derivative of $f$ evaluated on $\gamma=0$ is:
\[\left.\frac{df}{d\gamma}\right|_{\gamma=0}=-\frac{S}{2b^{3/2}}\frac{1-b\sin^2\psi}{\sqrt{a+b\sin^2\psi}\sin^3\psi}.\]
After integration, we obtain:
\[\left.\frac{dF}{d\gamma}\right|_{\gamma=0}=\left[\frac{S(a-b-2ab)}{4(ab)^{3/2}}\ln\left(\frac{\sqrt{a}\cos\psi+\sqrt{a+b\sin^2\psi}}{\sin\psi}\right)+\frac{S}{4}\frac{\sqrt{a+b\sin^2\psi}\cos\psi}{b^{3/2}a\sin^2\psi} \right]_{\epsilon_0}^{\frac{\pi}{2}}.\]
Note that the condition $\epsilon_0\ll 1$ implies that the terms of the order of $1/\epsilon_0^2$ are dominant in the right-hand side of this equation. An asymptotic expansion of this expression gives:
\begin{equation}
\left.\frac{\partial F}{\partial \gamma}\right|_{\gamma=0}=-\frac{S}{4b^{3/2}\sqrt{a}\epsilon_0^2}+O(\frac{1}{\epsilon_0}).
\end{equation}
Finally, we can compute $\partial \epsilon/\partial \gamma|_{\gamma=0}$ by using Eq.~(\ref{eq_depsdgamma_form}):
\begin{equation}
\left.\frac{\partial\epsilon}{\partial\gamma}\right|_{\gamma=0}=-\frac{1}{4b\epsilon_0}+O(1).
\end{equation}
To summarize, the angle $\epsilon$ can be expressed as:
\begin{equation}
\epsilon(\gamma)=\epsilon_0-\frac{\gamma}{4b\epsilon_0}+O(\gamma^2),
\end{equation}
where $\epsilon_0$ is given by equation~(\ref{eps0}),
and $O(\gamma^2)$ is a function depending on $\epsilon_0$, such that $O(\gamma^2)/\gamma^2$ tends to a constant (depending on $\epsilon_0$), as $\gamma$ goes to zero.

\section*{Acknowledgments}
D. Sugny acknowledges support from the ANR-DFG research program Explosys (ANR-14-CE35-0013-01). This work has been done with the support of the Technical University of Munich, Institute for Advanced Study, funded by the German Excellence Initiative and the European Union Seventh Framework Programme under grant agreement 291763.



\begin{thebibliography}{99}

\bibitem{Golstein50} H. Goldstein, \emph{Classical Mechanics} (Addison-Wesley, Reading, MA, 1950).

\bibitem{efstathiou} K. Efstathiou, \emph{Metamorphoses of Hamiltonian Systems with Symmetry}, Lecture Notes in Mathematics Series-LNM 1864 (Springer-Verlag, Heidelberg, 2004)

\bibitem{landau} L. D. Landau and E. M. Lifshitz, \emph{Mechanics} (Pergamon Press, Oxford, 1960).

\bibitem{arnold} V. I. Arnol\'d, \emph{Mathematical Methods of Classical Mechanics} (Springer-Verlag, New York, 1989).

\bibitem{cushman} R. H. Cushman and L. Bates, \emph{Global Aspects of Classical Integrable Systems} (Birkhauser, Basel, 1997).

\bibitem{fomenko} A. V. Bolsinov and A. T. Fomenko, \emph{Integrable Hamiltonian systems: geometry, topology, classification} (Boca Raton, FL: Chapman \& Hall/CRC, 2004)

\bibitem{hannay} S. Golin, A. Knauf and S. Marmi, Comm. Math. Phys. 123, 95 (1989)

\bibitem{Duistermaat} J. J. Duistermaat, Commun. Pure Appl. Math. 33, 687 (1980)

\bibitem{montgomery} R. Montgomery, Am. J. Phys. 59, 394 (1991)

\bibitem{levi} M. Levi, Arch. Rational Mech. Anal. 122, 213 (1993)

\bibitem{natario} J. Natario, J. Geom. Mech. 2, 113 (2010)

\bibitem{MSA91} M. S. Ashbaugh, C. C. Chiconc and R. H. Cushman, J. Dyn. Diff. Eq. 3, 67 (1991).

\bibitem{video} See supplemental material for videos about the tennis racket effect.

\bibitem{standard} A standard tennis racket is the Wilson T-2000. For this racket, the principal moments of inertia are  $I_1=0.121\times 10^{-2}kg.m^2$, $I_2=1.638\times 10^{-2}kg.m^2$, and $I_3=1.748\times 10^{-2} kg.m^2$. This leads to the values of the parameters: $a= 12.5372,\; b=0.0629$.

\bibitem{abramowitz} M. Abramowitz and I. A. Stegun, \emph{Handbook of Mathematical Functions}, Applied Mathematical Series, 55 (National Bureau of Standards, Washington, 1964)

\bibitem{nesbit} S. M. Nesbit, M. Elzinga, C. Herchenroder and M. Serrano, J. Sports Sci. Med. 5, 304 (2007)

\bibitem{landauQ} L. D. Landau and E. M. Lifshitz, \emph{Quantum Mechanics} (Pergamon Press, Oxford, 1973).

\bibitem{rouzee} A. Rouz\'ee, S. Gu\'erin, O. Faucher and B. Lavorel, Phys. Rev. A 77, 043412 (2008)

\bibitem{lee} K. F. Lee, D. M. Villeneuve, P. B. Corkum, A. Stolow and J. G. Underwood, Phys. Rev. Lett. 97, 173001 (2006)

\bibitem{levitt} M. H. Levitt, \emph{Spin Dynamics: Basics of Nuclear Magnetic Resonance} (Wiley, New York, 2008).

\bibitem{cat} S. J. Glaser, U. Boscain, T. Calarco, C. P. Koch, W. Kockenberger, I. Kuprov, B. Luy, S. Schirmer, T. Schulte-Herbruggen, D. Sugny and F. Wilhelm, Eur. Phys. J. D 69, 279 (2015).

\bibitem{lapert} M. Lapert, Y. Zhang, M. Braun, S. J. Glaser, and D. Sugny, Phys. Rev. Lett. 104, 083001 (2010).

\bibitem{vandamme} L. Van Damme, R. Zeier, S. J. Glaser and D. Sugny, Phys. Rev. A 90, 013409 (2014)

\bibitem{nimbalkar} M. Nimbalkar, R. Zeier, J. L. Neves, S. B. Elavarasi, H. Yuan, N. Khaneja, K. Dorai and S. J. Glaser, Phys. Rev. A 85, 012325 (2012)

\bibitem{brody1} H. Brody, Am. J. Phys. 47, 482 (1979)

\bibitem{brody2} H. Brody, Phys. Teach. 23, 213 (1985)

\bibitem{rogowski} I. Rogowski, T. Creveaux, L. Cheze, P. Mace and R. Dumas, Plos one 9, 104785 (2014)
\end{thebibliography}
\end{document}